\newif\iftwocol
\twocoltrue

\iftwocol
\documentclass[journal,10pt,final,twocolumn]{IEEEtran}
\else
\documentclass[journal,11pt,draftclsnofoot,onecolumn]{IEEEtran}
\fi

\usepackage{amsmath,graphicx}
\usepackage{amsfonts, amssymb, ifthen}
\usepackage[lined,linesnumbered]{algorithm2e}
\usepackage[utf8]{inputenc}
\usepackage{xcolor}
\usepackage{booktabs}

\newcommand{\rev}[1]{{\color{black}#1}} 
\newcommand{\brev}{\color{black}} 
\newcommand{\erev}{\color{black}}

\def\integers{\ensuremath{\mathbb{Z}}}
\def\MAE{\ensuremath{\mathrm{MAE}}}

\newcommand{\maxCodeWordLength}{\tau}
\newcommand{\GolombForget}{F}
\newcommand{\blockLength}{B}
\newcommand{\CodeLength}[3]{C_{#1,#2}(#3)}
\newcommand{\AvgCodeLength}[3]{\bar{C}_{#1,#2}(#3)}
\newcommand{\DMST}[1]{\hat{T}\left(G_{#1}\right)}
\newcommand{\stopn}{n_s}

\newcommand{\sample}[2]{x_{#1}(#2)}
\newcommand{\qsample}[2]{\tilde{x}_{#1}(#2)}
\newcommand{\samples}[2]{\mathbf{x}_{#1}(#2)}
\newcommand{\samplePred}[2]{\hat{x}_{#1}(#2)}

\newcommand{\samplePredP}[2]{\hat{x}^p_{#1}(#2)}
\newcommand{\samplePredOrderSeq}[3]{\dot{x}^{#3}_{#1}(#2)}
\newcommand{\samplePredPSeq}[2]{\samplePredOrderSeq{#1}{#2}{p}}

\newcommand{\errorNotation}{\epsilon}
\newcommand{\errorSingleCh}[1]{\errorNotation(#1)}
\newcommand{\error}[2]{\errorNotation_{#1}(#2)}
\newcommand{\qerror}[2]{\tilde{\epsilon}_{#1}(#2)}
\newcommand{\totalErrorP}[2]{E^p_{#1}(#2)}
\newcommand{\totalErrorPSeq}[2]{\mathcal{E}^p_{#1}(#2)}
\newcommand{\neighbor}[1]{\ell}
\newcommand{\rootch}{r}

\newcommand{\lsParams}[3]{\mathbf{a}_{#1}^{#3}(#2)}
\newcommand{\round}[1]{\left\lfloor #1 \right\rceil}
\newcommand{\Alphabet}{\mathcal{X}}

\newcommand{\ourmvar}{\mbox{$\text{MVAR}_2^n$}}

\usepackage{subcaption}
\usepackage{balance}

\newcommand{\squeeze}[1][-4pt]{
\addtolength{\abovedisplayskip}{#1}
\addtolength{\belowdisplayskip}{#1}
\addtolength{\belowdisplayshortskip}{#1}
\addtolength{\abovedisplayshortskip}{#1}
}

\begin{titlepage}
\title{Efficient sequential compression of multi-channel biomedical signals}

\author{Ignacio Capurro, Federico Lecumberry,~\IEEEmembership{Member, IEEE}, Álvaro Martín, Ignacio Ramírez,~\IEEEmembership{Member, IEEE}, Eugenio Rovira and Gadiel~Seroussi,~\IEEEmembership{Fellow, IEEE}
\thanks{This work was funded by CSIC, Universidad de la República.}
\thanks{Preliminary results of this work were presented at the 2014 European Signal Processing Conference (EUSIPCO 2014), Lisbon, Portugal, 2014}
\thanks{Authors are with Universidad de la Rep\'{u}blica, Montevideo, Uruguay (e-mails: \{icapurro, fefo, almartin, nacho, exavier\}@fing.edu.uy, and gseroussi@ieee.org)}
}

\end{titlepage}

\begin{document}
\maketitle
\begin{abstract}
  This work proposes lossless and near-lossless compression algorithms
  for multi-channel biomedical signals.  The algorithms are sequential
  and efficient, which makes them suitable for low-latency and
  low-power signal transmission applications.  We make use of
  information theory and signal processing tools (such as universal
  coding, universal prediction, and fast online implementations of
  multivariate recursive least squares), combined with simple methods
  to exploit spatial as well as temporal redundancies typically
  present in biomedical signals.  The algorithms are tested with
  publicly available electroencephalogram and electrocardiogram
  databases, surpassing in all cases the current state of the art in
  near-lossless and lossless compression ratios.

\end{abstract}
\section{Introduction}
\label{sec:intro}

Data compression is of paramount importance when dealing with biomedical data sources that produce
large amounts of data, due to the potentially large savings in storage and/or transmission costs.
%
Some medical applications involve real-time monitoring of patients and individuals in
their everyday activities (not confined to a bed or chair).
In such contexts, wireless and self-powered acquisition devices are 
desirable, 
imposing severe power consumption restrictions that call for efficient 
bandwidth use and simple embedded logic.

The requirements imposed by these applications, namely,
low-power, and efficient on-line transmission of the data,
naturally lead to a requirement of low-complexity, and low-latency,
compression algorithms. Also, as it is typical in medical applications, data acquired for clinical
purposes is often required to be transmitted
and/or stored without or at worst with very small distortion with respect to the data that was acquired by the
sensors. This in turn leads to a need for lossless or near-lossless algorithms, where every decoded
sample is guaranteed to differ by up to a preestablished bound from the 
original sample (in the lossless case, no differences are allowed).
\brev





\subsection{Background on biomedical signal compression}

Most lossless biomedical signal compression methods (both single- and 
multi-channel) are based on a
predictive stage for removing temporal and/or spatial correlations. This produces a predicted
signal, which is then subtracted from the original signal to obtain a prediction error that is
encoded losslessly~\cite{Koski199723,antoniol97tbe,arnavut09icsccw,srinivasan11bspc,Memon1999}.
Among prediction methods, typical choices include linear
predictors~\cite{Koski199723,arnavut09icsccw}, and neural networks~\cite{antoniol97tbe}. After
correlation is (hopefully) removed by the prediction stage, residuals are encoded according to some
statistical model. Typical choices include arithmetic, Huffman, and Golomb 
encoding. An example of a
non-predictive scheme, based on a lossless variant of the typical 
transform-based methods used in
lossy compression for temporal decorrelation, is given 
in~\cite{wongsawat06iscas}.

When considering the multi-channel case, a spatial decorrelation stage is 
generally included to account for
inter-channel redundancy.  In this case, most lossless and near-lossless algorithms found in the
literature resort to transform-based approaches to remove spatial redundancy. This
includes~\cite{wongsawat06iscas}, and the works~\cite{srinivasan13bhi,dauwels13bhi}, where a lossy
transform is used for simultaneous spatio-temporal decorrelation, and the residuals are encoded
losslessly or near-losslessly. In~\cite{srinivasan13bhi} this is done using a wavelet transform,
whereas ~\cite{dauwels13bhi} uses a PCA-like decomposition. As an example of a non-transform based
method, the work~\cite{liu02icsp} decorrelates a given channel by subtracting, from each of its
samples, a fixed linear combination of the samples from neighboring electrodes corresponding to the
same time slot. Another example is the MPEG-4 audio lossless coding standard~\cite{ALS} (ALS), which
has also been applied for biomedical signal compression~\cite{bioALS}. In this case, a linear
prediction error is obtained for each channel, and the inter-channel correlation is exploited by
subtracting, from each prediction error in a target channel, a linear combination of prediction
errors of a reference channel. The signal is divided into blocks and for each block, two passes are
performed through the data. In the first pass, the pairs of target-reference channel and the
coefficients for intra and inter channel linear predictions are obtained and described to the
decoder; the data itself is encoded in the second pass.

In relation to the objectives posed for this work, we note that algorithms
such as those described in
~\cite{wongsawat06iscas,srinivasan11bspc,srinivasan13bhi,dauwels13bhi}
require a number of operations that is superlinear in the number of
channels.
%
%
Some methods,
including~\cite{wongsawat06iscas,arnavut09icsccw,srinivasan11bspc,srinivasan13bhi,dauwels13bhi, ALS}
also have the drawback of having to perform more than one pass over the input 
data. 
Arithmetic coders such as those used in
\cite{arnavut09icsccw,srinivasan11bspc,srinivasan13bhi,dauwels13bhi} are
computationally more expensive than, for example, a Golomb 
coder~\cite{Golomb:66},
which is extremely simple and thus popular in low-power embedded
systems. Finally, methods such as
\cite{srinivasan11bspc,srinivasan13bhi,dauwels13bhi} split the signal into
an approximate transform and a residual, and perform an exhaustive search
among several candidate splittings, where each candidate splitting is
tentatively encoded, thus further increasing the coding time.

\subsection{Contribution}

In the present work we propose a sequential, low-latency,
low-complexity, lossless/near-lossless compression algorithm. The algorithm
uses a statistical model of the signals, designed with the goal of exploiting,
simultaneously, the potential redundancy appearing across samples at different
times (temporal redundancy) and the redundancy among samples obtained from different channels
during the same sampling period (spatial redundancy). The design relies on 
well-established theoretical tools from  universal 
compression~\cite{MSW2000,Riss84} and
prediction~\cite{singer99tsp},
combined with advanced signal processing tools~\cite{Glentis1995a}, to define
a sequential predictive coding scheme that, to the best of our knowledge, has 
not been introduced before.
The results are backed by extensive experimentation on
publicly available electroencephalogram (EEG) and electrocardiogram (ECG) 
databases, showing the best lossless and near-lossless
compression ratios for these databases in the state of the art (in the  
near-lossless case, compression ratios are compared for the same distortion 
level under a well defined metric).

The execution time of the proposed algorithm is
linear in both the number of channels and of time samples, requiring an amount
of memory that is linear in the number of channels.  The algorithm relies on
the observation that inter-channel redundancy can be effectively reduced by
jointly coding channels generated by sensors that are physically close to
each other. The specific statistical model derived from this observation is
detailed in Section~\ref{sec:modeling}, and an encoding algorithm based on
this model is defined in Section~\ref{sec:proposal}. In the usual scenario
in which the sensor positions of the acquisition system are known, the
proposed algorithm, referred to as Algorithm~\ref{alg:Code}, defines a
simple and efficient joint coding scheme to account for inter-channel
redundancy. If these positions are unknown, we propose, in
Section~\ref{ssec:unknown}, an alternative method, implemented in
Algorithm~\ref{alg:Code2}, to sequentially derive an adaptive joint coding
scheme from the signal data that is being compressed. This is accomplished
by collecting certain statistics simultaneously with the compression of a
segment of initial samples. During this period, which is usually short (in
general between 1000 and 2000 vector samples), the execution time and memory
requirements are quadratic in the number of channels.  In both scenarios, the
proposed algorithms are sequential (the samples are encoded as soon as they are received).

In Section~\ref{sec:results} we provide experimental evidence on the
compression performance of the proposed algorithms, showing compression 
ratios that surpass the published 
state-of-the-art~\cite{srinivasan13bhi,dauwels13bhi,ALS}. 
The compression ratios obtained with
Algorithms~\ref{alg:Code}
and~\ref{alg:Code2} are similar, showing that there is
no significant compression performance loss for lack of prior knowledge of 
the device sensor positions. Moreover, Algorithm~\ref{alg:Code2} achieves 
slightly better compression ratios in some cases. Final conclusions are 
discussed in Section~\ref{sec:concl}.

\subsection{Summary of the contribution}

In summary, our contribution consists of two algorithms, whose properties
are summarized below:

\begin{itemize}
\item \textbf{Sequential/online.} Data is processed and transmitted as it
  arrives. Note that, of the algorithms that report the current best
  compression rates in the literature, only~\cite{ALS} can be applied
  online. The rest require multiple passes over the data.

\item \textbf{Low latency.} Since the algorithms are sequential, the only 
latency
  involved is the CPU time required to process a scalar sample.
  The latency of the only competing
  method which is online~\cite{ALS} is $2048$ time samples, which represents a
  minimum of two seconds if operating at $1\,\text{kHz}$.

\item \textbf{Compression performance.} Both algorithms surpass the current 
state of the art in
  lossless compression algorithms, and also in near-lossless compression algorithms
  for the maximum absolute error (M*AE) distortion measure.

\item \textbf{Low complexity.} This is especially true for Algorithm~1,
  which requires a fixed, small number of computations per scalar sample.
  The complexity of Algorithm~2 is
  quadratic in the number of channels during a small number of initial
  samples (approximately $1000$, see Table~\ref{tab:stop} in
  Section~\ref{sec:results}), becoming identical to that of Algorithm~1 afterwards.

\end{itemize}


\erev
\section{Statistical modeling of biomedical signals for predictive coding}
\label{sec:modeling}

\subsection{Biomedical signals}
An \emph{electroencephalogram (EEG)} is a signal obtained through a set of electrodes placed on the scalp of a person or animal.
Each electrode measures the electrical activity produced by the neurons in certain region of the brain,
generating a scalar signal that is usually referred to as a \emph{channel}.
EEGs are commoly used in some clinical diagnostic techniques, and also find 
applications in biology, medical research, and brain-computer interfaces. For 
most clinical applications, the electrodes are placed on the scalp following 
the
international 10–20 system~\cite{System1020}, or a superset of it when a higher spatial resolution is required.
Depending on the specific goal, an EEG can be comprised of up to hundreds of channels (for example, our experiments include cases with up to 118 channels).
Modern electroencephalographs produce discrete signals, with sampling rates typically ranging from 250Hz to 2kHz and sample resolutions between 12 and 16 bits per channel sample.

An \emph{electrocardiogram (ECG)} is a recording of the heart electrical 
activity through electrodes that are usually placed on the chest and limbs of 
a person. A \emph{lead} of an ECG is a direction along which the heart 
depolarization is measured. Each of these measures is determined by linearly 
combining the electrical potential difference between certain electrodes. A 
standard 12-lead ECG record~\cite{macfarlane2010comprehensive}, for example, 
is comprised of 12 leads named \emph{i, ii, iii, aVR, aVL, aVF}, $v_1, 
\ldots, v_6$, which are obtained from 10 electrodes. Lead \emph{i}, for 
example, is the potential difference registered by electrodes in the left and 
right arms. In the description of our algorithm in the sequel we will 
use the term \emph{channel} generically, with the understandong that it 
should be interpreted as \emph{lead} in the case of ECGs.


\subsection{Predictive coding}
We consider a discrete time $m$-channel signal, $m>1$. We denote by $\sample{i}{n}$ the $i-$th channel (scalar) sample at time instant $n$, $n\in\mathbb{N}$, and we refer to the vector $(\sample{1}{n}, \ldots, \sample{m}{n})$ as the \emph{vector sample} at time instant $n$. We assume that all scalar samples are quantized to integer values in a finite interval $\Alphabet$.

The lossless encoding proposed in this paper follows a predictive coding 
scheme, in which a \emph{prediction} $\samplePred{i}{n}$ is sequentially 
calculated for each sample $\sample{i}{n}$, and this sample is described by 
encoding the \emph{prediction error}, $\error{i}{n}\triangleq \sample{i}{n} - 
\samplePred{i}{n}$. The sequence of sample descriptions is \emph{causal}, 
i.e., the order in which the samples are described, and the definition of the 
prediction $\samplePred{i}{n}$, are such that the latter depends solely on 
samples that are described before sample $\sample{i}{n}$. Thus, a decoder can 
sequentially calculate $\samplePred{i}{n}$, decode $\error{i}{n}$, and add 
these values to reconstruct $\sample{i}{n}$. A near-lossless encoding is 
readily derived from this scheme by quantizing the prediction error 
$\error{i}{n}$ to a value $\qerror{i}{n}$ that satisfies 
$|\error{i}{n}-\qerror{i}{n}|\leq \delta$, for some preset parameter 
$\delta$. After adding $\qerror{i}{n}$ to $\samplePred{i}{n}$, the decoder 
obtains a sample approximation, $\qsample{i}{n}$, whose distance to 
$\sample{i}{n}$ is at most $\delta$. In this case, the prediction 
$\samplePred{i}{n}$ may depend on the approximations  $\qsample{i}{n'}$, $n' 
< n$, of previously described samples, but not on the exact sample values, 
which are not available to the decoder.

\subsection{Statistical modeling}

The aim of the prediction step is to produce a sequence of prediction errors that, according to some preestablished probabilistic model, exhibit typically a low empirical entropy, which is then exploited in a coding step to encode the data economically.
In our encoder we use an adaptive linear predictor.
We model prediction errors by a \emph{two-sided geometric distribution} (TSGD), for which we
empirically observe a good fitting to the tested data, and which can be efficiently encoded with
adaptive Golomb codes~\cite{Golomb:66, MSW2000}. \brev The TSGD is a discrete distribution defined over the integers $\integers$ as,
\[
P(x;\theta,d)=\frac{1-\theta}{\theta^{1-d}+\theta^{d}}\theta^{|x+d|}\,,
\]
where $0 < \theta < 1 $ is a scale parameter and $0\leq d <1/2$ is a bias parameter. We refer the
reader to \cite{MSW2000} for details on the online adaptation of the TSGD parameters and the
corresponding optimal Golomb code parameters. For the following discussion, 
it suffices to note
that, under the TSGD distribution, the empirical entropy of an error sequence, $\errorSingleCh{1}\,,\ldots\,,
\errorSingleCh{N}$, of length $N$, is roughly proportional to $\log_2 \MAE(\errorNotation)$, where MAE stands for \emph{Mean
Absolute Error} and is defined as,
\begin{equation}\mathrm{MAE}(\errorNotation) = \sum_{n=1}^{N}|\errorSingleCh{n}|.
\label{eq:mae}
\end{equation}
As defined, $\log_2 \MAE$ provides an approximate measure of the relative savings in terms of \emph{bits
per sample} (bps) obtained when using different prediction schemes. Below, we 
discuss the considerations that lead to our specific choice of prediction 
scheme. In this discussion, and in the sequel, we refer to various databases 
of digitized EEG 
recordings (e.g., DB1a, DB2a, DB2b, etc.) used in our experiments; detailed 
descriptions of these databases are provided in Section~\ref{sec:results}. 
\begin{table}
\begin{center}
\caption{\label{tab:pred-mae}Performance of different prediction schemes in terms of $\log_2 \MAE$
  on database DB1a. }
\begin{tabular}{lc}\hline
Model & $\log_2 \MAE$ \\\hline\hline
AR, order 3         &	1.883    \\
MVAR, order 3 	& 1.763	\\
MVAR, order 6 	& 1.474	\\
$\mbox{MVAR}_2$, order 3 & 1.854 \\
{\bf \ourmvar, order 3}	& {\bf 1.351} \\
\hline
\end{tabular}
\end{center}
\end{table}
\erev

In an (independent channel) \emph{autoregressive model} (AR) of \emph{order} 
$p$, $p\geq 1$, every sample $\sample{i}{n}$, $n>p$, is the result of adding 
independent and identically distributed noise to a \emph{linear prediction}
\begin{equation}\squeeze
  \samplePredP{i}{n}=\sum_{k=1}^p a_{i,k}\sample{i}{n-k}\,,\quad 1\leq i\leq m\,,
\label{eq:AR}
\end{equation}
where the real coefficients $a_{i,k}$ are \emph{model parameters}, which 
determine, for each channel $i$, the dependence of $\sample{i}{n}$ on 
previous samples of the same channel. The prediction in a \emph{multivariate 
autoregressive model} (MVAR)
for a sample from a channel $i$, $1\leq i\leq m$, is
\begin{equation}\squeeze
  \samplePredP{i}{n}=\sum_{j=1}^m\sum_{k=1}^p a_{i,j,k}\sample{j}{n-k}\,,
\label{eq:MVAR}
\end{equation}
where now the model is comprised of $pm^2$ parameters, $a_{i,j,k}$, which define, for each $i$, a linear combination of past
samples from \emph{all} channels $j$, $1\leq j\leq m$. Consequently, this model may potentially capture both time and space signal
correlation. Indeed, for EEG data, we experimentally observe that the \rev{MAE}, where model
parameters are obtained as the solution to a least squares minimization, is in general significantly
smaller for an MVAR model than an AR model. For instance, Table~\ref{tab:pred-mae} shows a potential
saving of $0.12$ bits per sample (bps) on average (over all files and 
channels of the database DB1a) by using an MVAR model of order 3 instead of 
an AR model of the same order.

Some EEG signals, however, consist of up to hundreds of channels and, therefore, the number of model
parameters in~(\ref{eq:MVAR}) may be very large. As a consequence, since these parameters are
generally unknown a priori, MVAR models may suffer from a high statistical 
\emph{model cost} \brev (i.e., the cost of either describing or adaptively 
learning 
the model parameters)~\cite{Riss84}\erev, which may offset in practice the 
potential code length savings \brev shown in Table~\ref{tab:pred-mae}. As a 
compromise, one could use an MVAR based on a subset of the channels. 
For example, Table~\ref{tab:pred-mae} shows results for a model, referred to 
as $\mbox{MVAR}_2$, in which the 
prediction $\samplePredP{i}{n}$ is a linear combination of the $p$ most 
recent past samples from just two channels, $i$, $\neighbor{i}$, where 
$\neighbor{i}$ is a channel whose recording electrode is physically close to 
that of channel $i$. The result for $\mbox{MVAR}_2$ in the table shows that 
adding the second channel to the 
predictor indeed reduces the MAE; however, the gains over an AR of the same 
order are modest. On the other hand, we observed that considerably more gains 
are obtained if, 
besides past samples from channels $i$, $\neighbor{i}$, we also 
use the sample at time instant $n$ of channel $\neighbor{i}$ to predict 
$\sample{i}{n}$ (assuming causality is maintained, as will be discussed 
below), i.e.,
\begin{equation}
  \samplePredP{i}{n}=\sum_{k=1}^p a_{i,k}\sample{i}{n-k}+\sum_{k=0}^p b_{i,k}\sample{\neighbor{i}}{n-k}\,,
\label{eq:linearPredPresent}
\end{equation}
$1\leq i, \neighbor{i}\leq m,\,\, i\neq \neighbor{i}$.  We refer to this 
scheme as \ourmvar. As seen on Table~\ref{tab:pred-mae}, 
\ourmvar, with $p=3$, surpasses the performance of an MVAR model of order $3$
by over $0.4$bps, and even that of an MVAR model of order $6$ by $0.1$bps, 
with a much smaller model cost. In light of these results, \ourmvar\ was 
adopted as the prediction scheme in our compression algorithm.
\erev

\section{Encoding}
\label{sec:proposal}

In this section we define the proposed encoding scheme. Since the sequence of
sample descriptions must be causal with respect to the predictor, not all
predictions $\samplePred{i}{n}$ can depend on a sample at time $n$. Hence, in
Subsection~\ref{sseq:knownPositions} we define an order of description \brev
that obeys the causality constraint, and also minimizes the sum of the
physical distances between electrodes of channels $i$, $\neighbor{i}$, where
$\samplePred{i}{n}$ depends on $\sample{\neighbor{i}}{n}$,
over all channels $i$ except the one whose sample is described first. The
prior assumption here is that the correlation between the signals of two
electrodes will tend to increase as their physical distance decreases. \erev
In Subsection~\ref{ssec:codingAlg} we present the encoding process in full detail and in Subsection~\ref{ssec:unknown} we generalize the encoding scheme to the case in which the electrode positions are unknown. Finally, in Subsection~\ref{ssec:nearlossless} we present a near-lossless variant of our encoder. Experimental results are deferred to Section~\ref{sec:results}.

\begin{figure}
\centering
  \includegraphics[width=0.7\columnwidth]{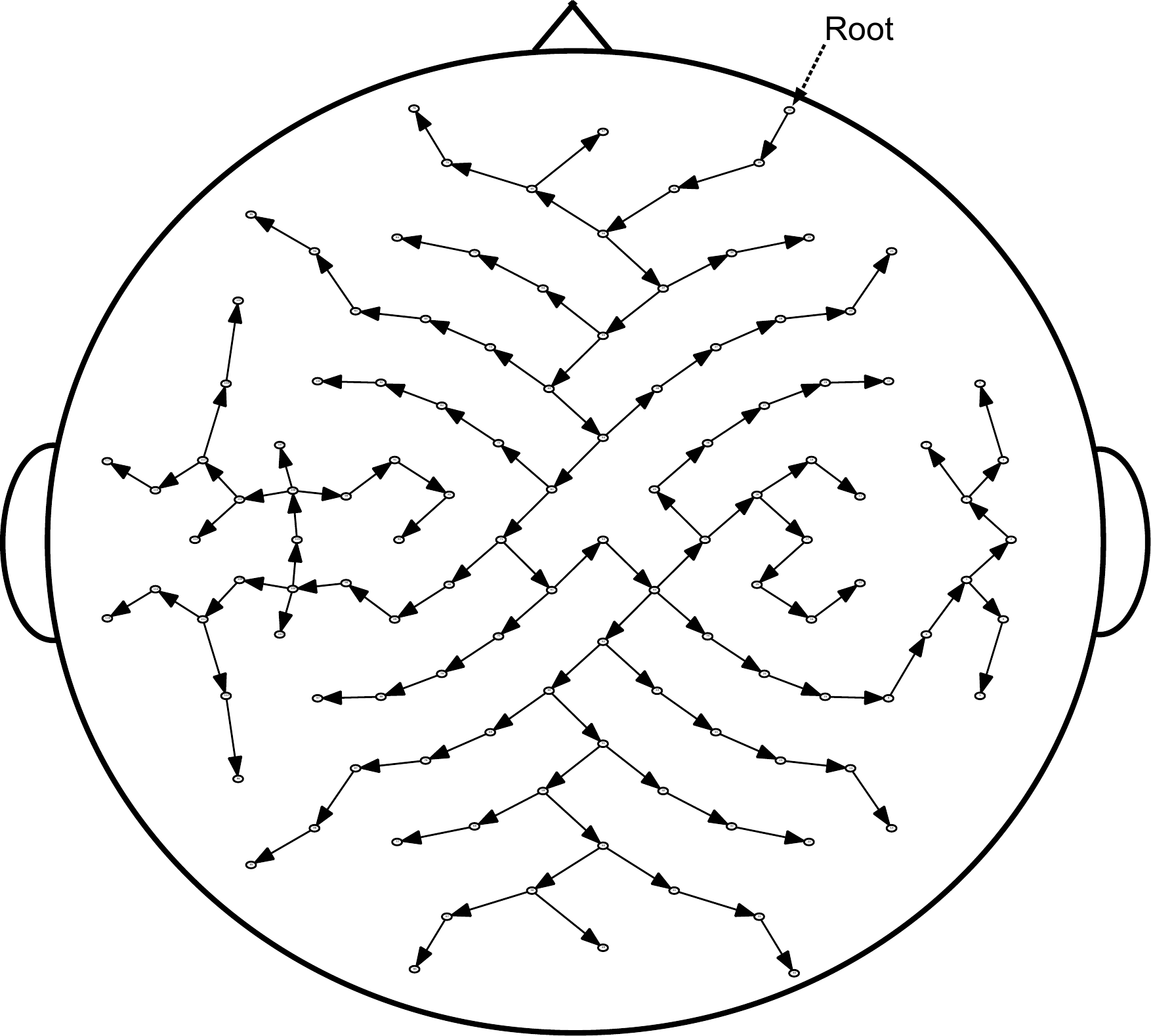}
  \caption{ \label{fig:tree} A graphical representation of the coding tree used in our experiments with EEG files from databases DB2a and DB2b (see Section~\ref{sec:results}).}
\end{figure}

\subsection{Definition of channel description order for known electrode positions}\label{sseq:knownPositions}
To define a channel description order we consider a tree, $T$, whose set of vertices is the set of channels, $\{1, \ldots, m\}$. We refer to $T$ as a \emph{coding tree}.
Specifically, in the context in which the electrode positions are known, we
let $T$ be a \emph{minimum spanning tree} \cite{MSTBoruvka,Kruskal1956} of the
complete graph whose set of vertices is $\{1,
\ldots, m\}$, and each edge $(i, j)$ is weighted with the \brev physical
\erev distance between electrodes of channels
$i,j$. In other words, the sum of the distances between electrodes of channels $i$, $j$, over all
edges $(i,j)$ of $T$, is minimal among all trees with vertices $\{1, \ldots,
m\}$. We distinguish an
arbitrary channel $\rootch$ as the \emph{root},\footnote{The specific selection of the root channel
  $\rootch$ did not have any significant impact on the results of our experiments\brev; for all databases
reported in Section~\ref{sec:results}, the difference between the best and worst root choices was
always less than $0.01$bps\erev.} and we let the edges of $T$ be oriented so
that there exists a (necessarily unique) directed path from $\rootch$ to
every other vertex of $T$. Since a tree has no cycles, the edges of $T$
induce a causal sequence of sample descriptions, for example, by arranging
the edges of $T$, $e_1, \ldots, e_{m-1}$, in a \brev \emph{breadth-first
traversal} \erev \cite{DFSearch} order. An example of a coding tree
used in our EEG compression experiments is shown in Figure~\ref{fig:tree}.
After describing a root channel sample, all other samples are described in the order in which their channel appear as the destination of an edge in the sequence $e_1, \ldots, e_{m-1}$. Notice that since $T$ depends on the acquisition system but not on the signal samples, this description order may be determined off-line. The sample $\sample{r}{n}$ is predicted based on samples of time up to $n-1$ of the channels $r$, $i$, where $(r,i)$ is the edge $e_1$; all other predictions, $\samplePred{i}{n}$, $i\neq\rootch$, depend on the sample at time $n$ of channel $\neighbor{i}$ and past samples of channels $\neighbor{i}$, $i$, where $(\neighbor{i},i)$ is an edge of $T$.

\begin{algorithm}[t]
\For{$n=1,2,\ldots$}{
Let $(r,i)$ be edge $e_1$ of $T$\\
$\samplePred{r}{n} = f_r(\samples{r}{n-1}, \samples{i}{n-1})$\label{line:predictRoot}\\
Encode $\error{r}{n}$\label{line:encodeRoot}\\

\For{$k=1, \ldots, m-1$}{
Let $(\neighbor{i},i)$ be edge $e_k$ of $T$\\
$\samplePred{i}{n} = f_i(\samples{i}{n-1}, \samples{\neighbor{i}}{n})$\label{line:predictCh}\\
Encode $\error{i}{n}$\label{line:encodeCh}\\
}
}
\caption{\label{alg:Code} Coding algorithm with fixed coding tree.
See sections~\ref{sseq:knownPositions} and~\ref{ssec:codingAlg} for notation and definitions. }
\end{algorithm}


\begin{figure*}
\begin{center}\includegraphics[width=0.7\textwidth]{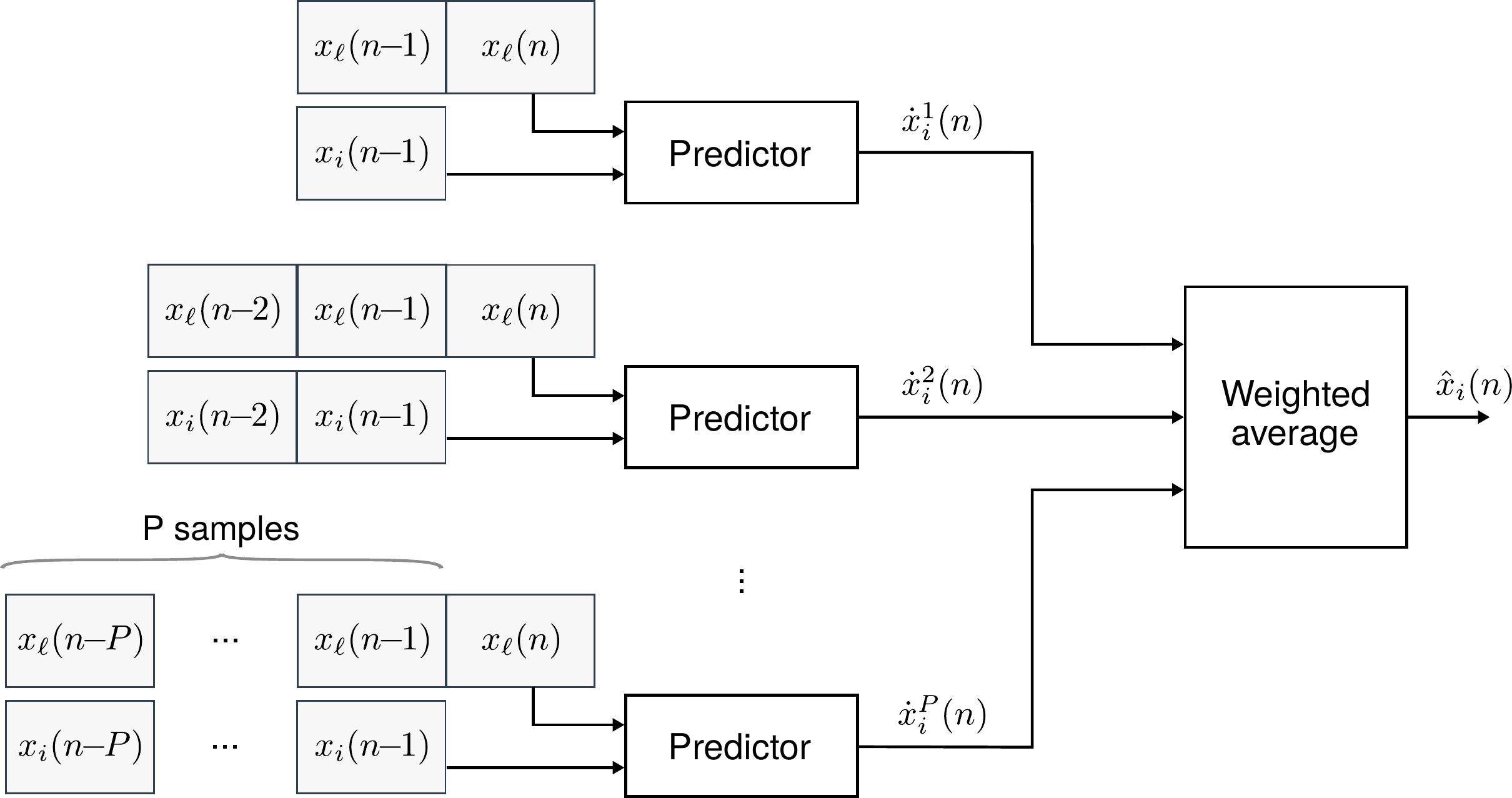}\end{center}
\caption{ \label{fig:weightedprediction} A representation of the weighted average predictor defined in~(\ref{eq:defPrediction}), for $i\neq \rootch$.}
\end{figure*}

\subsection{Coding algorithm}\label{ssec:codingAlg}
Algorithm~\ref{alg:Code} 
summarizes the proposed encoding scheme. We let $\samples{i}{n}=
\sample{i}{1}, \ldots, \sample{i}{n}$ denote the sequence of the first $n$
samples from channel $i$, and we let $f_i$ be an integer valued prediction
function to be defined.

We use adaptive Golomb codes~\cite{Golomb:66} for the encoding of
prediction errors in steps~4 
and~8. 
\brev Golomb codes are prefix-free codes on the integers,
characterized by very simple encoding and decoding operations
implemented with integer manipulations without the need for any
tables; they were proven optimal for geometric
distributions~\cite{Gall:75}, and, in appropriate combinations, also
for TSGDs~\cite{OptTSGD}. A Golomb code is characterized by a positive
integer parameter referred to as its \emph{order}.  To use Golomb
codes adaptively in our application, \erev an independent set of
prediction error statistics is maintained for each channel, \brev
namely the sum of absolute prediction errors and the number of encoded
samples\erev.  The statistics collected up to time $n-1$ determine the
order of a Golomb code, which is combined with a Rice
mapping~\cite{Rice} from integers to nonnegative integers to encode
the prediction error at time $n$.  \brev Both statistics are halved
every $\GolombForget$ samples to make the order of the Golomb code
more sensitive to recent error statistics and adapt quickly to changes
in the prediction performance\erev. Prediction errors are reduced
modulo $|\Alphabet|$ and long Golomb code words are escaped so that no
sample is encoded with more than a prescribed constant number of bits,
$\maxCodeWordLength$. This encoding of prediction errors is
essentially the same as the one used in~\cite{JPEG-LS}. In particular,
only Golomb code orders that are powers of two are used.

To complete the description of our encoder, we next define the
prediction functions, $f_i$, $1\leq i\leq m$, which are used in
steps~3 
and~7 
of Algorithm~\ref{alg:Code}. For a model order $p$, we let
$\lsParams{i}{n}{p}= \{a_{i,k}(n)$, $b_{i,k}(n)\}$ denote the set of
coefficient values, $a_{i,k}$, $b_{i,k}$, that, when substituted
into~(\ref{eq:linearPredPresent}), minimize the total weighted squared
prediction error up to time $n$
\begin{equation}\squeeze
  \totalErrorP{i}{n}=\sum_{j=1}^n\lambda^{n-j}\Big(\sample{i}{j}-\samplePredP{i}{j}\Big)^2\,,
  \label{eq:totalErrorP}
\end{equation}
where $\lambda$, $0<\lambda < 1$, is an exponential decay factor
parameter. \brev This parameter has the effect of preventing
$\totalErrorP{i}{n}$ from growing unboundedly with $n$, and of
assigning greater weight to more recent samples, which makes the
prediction algorithm adapt faster to changes in signal
statistics\erev.
A \emph{sequential linear predictor} of order $p$ uses the
coefficients $\lsParams{i}{n-1}{p}$ to predict the sample value at
time $n$ as\footnote{Notice that, compared
  to~(\ref{eq:linearPredPresent}), we use a different notation for the
  predictors in~(\ref{eq:linearPredPresentSeq}) as these will be
  combined to obtain the final predictor $\hat{x}$ used in
  Algorithm~\ref{alg:Code}.}
\begin{equation}
  \samplePredPSeq{i}{n}\!=\!\sum_{k=1}^p \!a_{i,k}(n\!-\!1)\sample{i}{n\!-\!k}+\sum_{k=0}^p b_{i,k}(n\!-\!1)\sample{\neighbor{i}}{n\!-\!k}\,,
\label{eq:linearPredPresentSeq}
\end{equation}
and, after having observed $\sample{i}{n}$, updates the set of
coefficients from $\lsParams{i}{n-1}{p}$ to $\lsParams{i}{n}{p}$ and
proceeds to the next sequential prediction.  This determines a total
weighted \emph{sequential absolute prediction error} defined as
\begin{equation}\squeeze
  \totalErrorPSeq{i}{n}=\sum_{j=1}^n\lambda^{n-j}\Big|\sample{i}{j}-\samplePredPSeq{i}{j}\Big|\,.
\label{eq:totalErrorPSeq}
\end{equation}
Notice that each prediction $\samplePredPSeq{i}{j}$
in~(\ref{eq:totalErrorPSeq}) is calculated with a set of model
parameters, $\lsParams{i}{j-1}{p}$, which only depends on samples that
are described before $\sample{i}{j}$ in
Algorithm~\ref{alg:Code}. These model parameters vary, in general,
with $j$ (cf.~(\ref{eq:totalErrorP})).

\brev
Various algorithms have been proposed to efficiently calculate
$\lsParams{i}{n}{p}$ from $\lsParams{i}{n-1}{p}$ simultaneously \emph{for
all} model orders $p$, up to a predefined maximum order $P$. We resort, in
particular, to a \emph{lattice algorithm} (see, e.g.,~\cite{Glentis1995a} and
references therein).  This calculation requires a constant number of scalar
operations for fixed $P$, which is of the same order (quadratic in $P$) as
the number of scalar operations that would be required by a conventional
least square minimization algorithm to compute the single set of model
parameters $\lsParams{i}{n}{P}$ for the largest model order $P$.
Also, we notice that using a lattice algorithm, the coefficients
$\lsParams{i}{n-1}{p}$ involved in the
definition~(\ref{eq:linearPredPresentSeq}) of $\samplePredPSeq{i}{n}$ can be
sequentially computed simultaneously for all $p$, $0\leq p\leq P$\erev.
Therefore, following~\cite{singer99tsp}, we do not fix nor estimate any specific model order but we instead average the predictions of all sequential linear predictors of order $p$, $0\leq p \leq P$, exponentially weighted by their prediction performance up to time $n-1$. Specifically, for $i\neq r$, we define
\begin{equation}
f_i(\samples{i}{n-1}, \samples{\neighbor{i}}{n}) = \round{\frac{1}{M}\sum_{p=0}^P\mu_p(n)\samplePredPSeq{i}{n}}\,,
\label{eq:defPrediction}
\end{equation}
where $\round{\cdot}$ denotes rounding to the nearest integer within the
quantization interval $\Alphabet$,
\begin{equation}
\mu_p(n) = \exp\{-\frac{1}{c} \totalErrorPSeq{i}{n-1}\}\,,
\label{eq:defweight}
\end{equation}
$M$ is a normalization factor that makes $\frac{\mu_p(n)}{M}$ sum up to unity with $p$, $\totalErrorPSeq{i}{n-1}$ is defined in~(\ref{eq:totalErrorPSeq}), and $c$ is a constant that depends on $\Alphabet$~\cite{singer99tsp}. If the weights $\mu_p$ are exponential functions of the sequential \emph{squared} prediction error, it is shown in~\cite{singer99tsp}  that the per-sample normalized squared prediction error of this predictor is asymptotically as small as the minimum normalized sequential squared prediction error among all linear predictors of order up to $P$. In our experiments, the compression ratio is systematically improved if the weights are defined instead as exponential functions of the sequential \emph{absolute} prediction error as in~(\ref{eq:defweight}). Figure~\ref{fig:weightedprediction} shows a schematic representation of the predictor $f_i$ defined in~(\ref{eq:defPrediction}). The definition of $f_r$ is analogous, removing the terms corresponding to $k=0$ from~(\ref{eq:linearPredPresent}) and~(\ref{eq:linearPredPresentSeq}), and letting the summation index $p$ in~(\ref{eq:defPrediction}) take values in the range $1\leq p\leq P+1$.

\begin{figure}
\centering
  \includegraphics[width=0.45\textwidth]{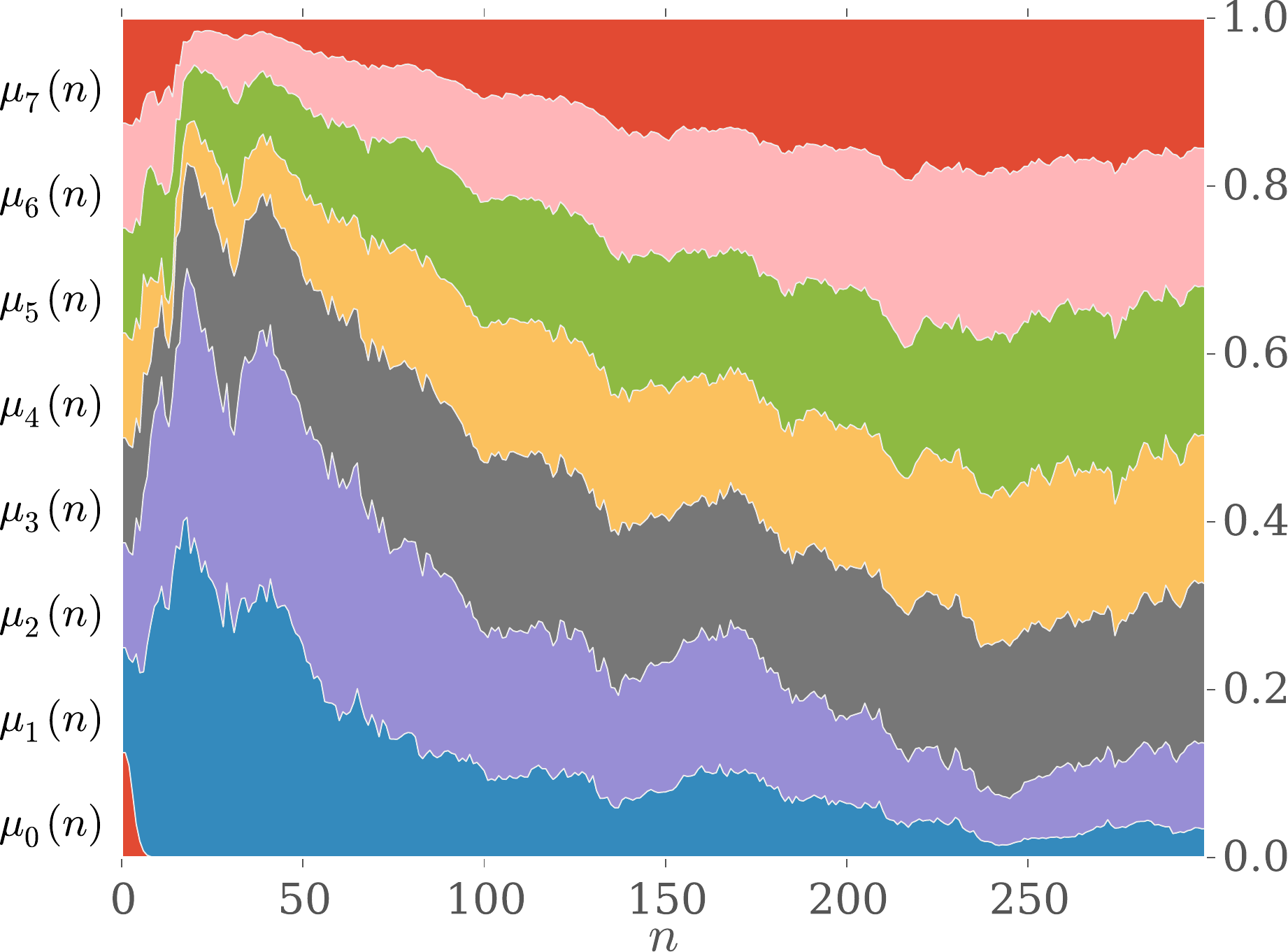}
  \caption{ \label{fig:weights} Stacked plot of the weights $\mu_p(n)$ for
  the first 300 samples of a 160Hz EEG channel from database DB1a.} %
\end{figure}
\begin{figure}
  \includegraphics[width=0.45\textwidth]{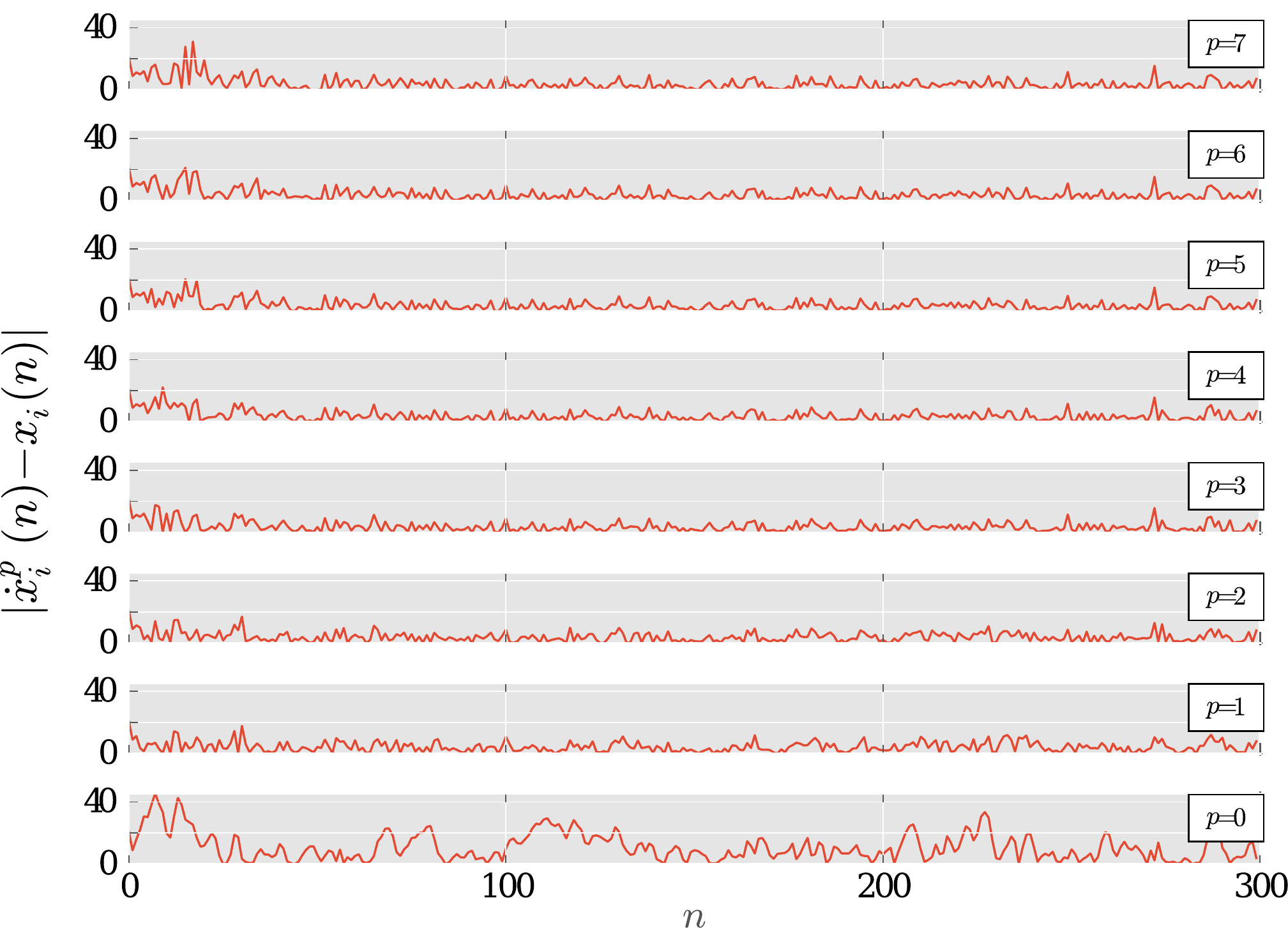}
  \caption{ \label{fig:errors} Absolute prediction error for the first 300 samples of the same EEG channel of Figure~\ref{fig:weights}.}
\end{figure}
Figure~\ref{fig:weights} shows the evolution of the weights $\mu_p(n)$,
$0\leq p \leq P$, during the initial segment of an EEG channel signal taken
from database DB1a. Small model orders, which adapt fast, receive high
weights for the very first samples. As $n$ increases, the performance of
large order models improves, and these orders gain weight.
Figure~\ref{fig:errors} shows the absolute prediction errors for the same EEG
channel.

The overall encoding and decoding time complexity of the algorithm is linear in the number of encoded samples. Indeed, a Golomb encoding over a finite alphabet requires $O(1)$ operations and, since the set of predictions $\samplePredPSeq{i}{n}$, $0\leq p\leq P$, can be recursively calculated executing $O(1)$ scalar operations per sample~\cite{Glentis1995a}, the sequential computation of $f_i$ requires $O(1)$ operations per sample. Regarding memory requirements, since each predictor and Golomb encoder requires a constant number of samples and statistics, the overall memory complexity of a fixed arithmetic precision implementation of the proposed encoder is $O(m)$.

\subsection{Definition of channel description order for unknown electrode positions}
\label{ssec:unknown}
\brev When the electrode positions are unknown, we derive the coding tree $T$
from the signal itself\erev.  To this end, we define an \emph{initial tree},
$T_0$, with an arbitrary root channel $r$ and a set of directed edges $\{(r,
i): 1 \leq i \leq m, i\neq r\}$ (a ``star'' tree). The first $\blockLength$
vector samples, $(\sample{1}{n}, \ldots, \sample{m}{n})$, $1\leq n\leq
\blockLength$, are encoded with Algorithm~\ref{alg:Code} setting $T=T_0$,
where $\blockLength$ is a fixed block length. We update $T$ every
$\blockLength$ vector samples until we reach a stoping time, $\stopn$, to be
defined. For each pair of channels $(\neighbor{i}, i)$, $i\neq \neighbor{i}$,
$i\neq r$,  we calculate the prediction $f_i(\samples{i}{n-1},
\samples{\neighbor{i}}{n})$ of $\sample{i}{n}$, defined
in~(\ref{eq:defPrediction}), and we determine the code length,
$\CodeLength{\neighbor{i}}{i}{n}$, of encoding the prediction error
$\sample{i}{n} - f_i(\samples{i}{n-1}, \samples{\neighbor{i}}{n})$ for all
$n$, $1\leq n \leq \stopn$. Notice that only the predictions
$f_i(\samples{i}{n-1}, \samples{\neighbor{i}}{n})$ such that $(\neighbor{i},
i)$ is an edge of $T$ are used for the actual encoding; the remaining
predictions and code lengths are calculated for statistical purposes with the
aim of determining a coding tree for the next block of samples.

We define a directed graph $G_n$ with a set of vertices $\{1, \ldots, m\}$ and a set of edges $\{(\neighbor{i}, i): i\neq \neighbor{i}, i\neq r\}$, where each edge $(\neighbor{i}, i)$ is weighted with the average code length $\AvgCodeLength{\neighbor{i}}{i}{n} = \frac{1}{n}\sum_{i = 1}^{n}\CodeLength{\neighbor{i}}{i}{n}$. Let $\DMST{n}$ be a directed tree with the same vertices as $G_n$, root $r$, and a subset of the edges of $G_n$ with minimum weight sum, i.e., $\DMST{n}$ is a minimum spanning tree of the directed graph $G_n$. Notice that, by the definition of $G_n$, setting  $T=\DMST{n}$ in Algorithm~\ref{alg:Code} yields the shortest code length for the encoding of the first $n$ vector samples among all possible choices of a tree $T$ with root $r$. However, to maintain sequentiality, $T$ can only depend on samples that have already been encoded. Therefore, for each $n$ that is multiple of the block length $\blockLength$, we set $T=\DMST{n}$ and use this coding tree to encode the next block, $(\sample{1}{i}, \ldots, \sample{m}{i})$, $n< i \leq n+\blockLength$. This sequential update of $T$ continues until a stopping condition is reached at some time $\stopn$; all subsequent samples are encoded with the same tree $\DMST{\stopn}$.

In our experiments, as detailed in Section~\ref{sec:results}, stopping when
the compression stabilizes yields small values of $\stopn$ and similar
compression ratios as Algorithm~\ref{alg:Code} with a fixed tree as defined
in Section~\ref{sseq:knownPositions}. Specifically, for the $i$-th block of
samples, $i > 0$, let $c_i$ be the sum of the edge weights of the tree
$\DMST{i\blockLength}$, i.e., $c_i$ is the average code length that would be
obtained for the first $i\blockLength$ vector samples with
Algorithm~\ref{alg:Code} using the coding tree determined upon encoding the
$i$-th block of samples. We also define $\Delta_i = |c_i - c_{i-1}|$, $i>1$,
and we let $\bar{\Delta}_i$ be the arithmetic mean of the last $V$ values of
$\Delta$, i.e., $\bar{\Delta}_i = V^{-1}\sum_{k=0}^{V-1}\Delta_{i-k}$, where
$V$ is some small constant and $i>V$. We define $\stopn$ as
\begin{equation}\label{eq:stopping}
  \stopn=\min\{N_s\} \cup \{i\blockLength: i > V, \bar{\Delta}_i < \gamma c_i\}\,,
\end{equation}
where the constant $N_s$ establishes a maximum for $\stopn$, and $\gamma$ is a constant.

\begin{algorithm}[h]
Set $T=T_0$ \\
Initialize $G_0$ with all edge weights equal to zero\\
Set $update$ = true\\
\For{$n=1,2,\ldots$}{
    Encode $(\sample{1}{n}, \ldots, \sample{m}{n})$ as in Algorithm~\ref{alg:Code} \\
    \If{$update$}{\label{line:ifupdate}
        Compute $G_n$ from $G_{n-1}$ and $\CodeLength{\neighbor{i}}{i}{n}, i\neq \neighbor{i}, i\neq r$\label{line:Gupdate}\\
        \If {$n$ is multiple of $\blockLength$}{
            Set $T=\DMST{n}$\label{line:Tupdate}\\
            \If {Stopping condition is true}{Set $update$ = false} \label{line:stop}
        }
    }
}
\caption{Coding algorithm with periodic updates of $T$. See Section~\ref{ssec:unknown} for notation and definitions.}\label{alg:Code2}
\end{algorithm}

The proposed encoding is presented in Algorithm~2. 
Step~7 
of Algorithm~2 
clearly requires $O(m^2)$ operations and $O(m^2)$ memory space.
Step~9 
also requires $O(m^2)$ operations and memory space using efficient
minimum spanning tree construction algorithms over directed
graphs~\cite{Tarjan, CameriniFM79, DMST}. Thus, compared to
Algorithm~\ref{alg:Code}, whose time and space complexity depend
linearly on $m$, steps! 7-12 
of Algorithm~\ref{alg:Code2} require an additional number of
operations and memory space that are quadratic in $m$. These steps,
however, are only executed until the stopping condition is true, which
in practice is usually a relatively small number of operations (see
Section~\ref{sec:results}).

\subsection{Near-lossless encoding}\label{ssec:nearlossless}
In a near-lossless setting, steps~4 
and~8 
of Algorithm~1 
encode a quantized version, $\qerror{i}{n}$, of each prediction error,
$\error{i}{n}$, defined as
\begin{equation}
  \qerror{i}{n} = \mbox{sign}(\error{i}{n})\left\lfloor\frac{|\error{i}{n}| + \delta}{2\delta + 1}\right\rfloor\,,
\end{equation}
where $\lfloor z \rfloor$ denotes the largest integer not exceeding $z$. This quantization guarantees that the reconstructed value, $\qsample{i}{n}\triangleq \samplePred{i}{n} + \qerror{i}{n}(2\delta+1)$, differs by up to $\delta$ from $\sample{i}{n}$.
All model parameters and predictions are calculated with $\qsample{i}{n}$ in
lieu of $\sample{i}{n}$, \brev on both the encoder and the decoder side.
Thus, the encoder and the decoder calculate exactly the same prediction for
each sample, and the distortion originated by the quantization of prediction
errors remains bounded in magnitude by $\delta$ (in particular, it does not
accumulate over time\erev). The code lengths
$\CodeLength{\neighbor{i}}{i}{n}$ in Algorithm~\ref{alg:Code2} are also
calculated for quantized versions of the prediction errors.

\section{Experiments, results and discussion}
\label{sec:results}

\subsection{Datasets}


We evaluate our algorithms by running lossless ($\delta=0$) and near-lossless
($\delta > 0$) compression experiments over the files of
several publicly available databases, described below. \brev In the
desciption, bps stands for ``bits per scalar sample''.\erev
\begin{itemize}
  \item DB1a and DB1b~\cite{PhysioNet,BCI2000} (BCI2000 instrumentation system): 64-channel, 160Hz, 12bps  EEG of 109 subjects using the BCI2000 system. The database consists of 1308 2-minute recordings of subjects performing a motor imagery task (DB1a), and 218 1-minute calibration recordings (DB1b).
  \item DB2a and DB2b~\cite{BCI3Berlin} (BCI Competition III,\footnote{http://bbci.de/competition/iii/} data set IV): 118-channel, 1000Hz, 16bps EEG of 5 subjects performing motor imagery tasks (DB2a). The average duration over the 8 recordings of the database is 39 minutes, with a minimum of almost 13 minutes and a maximum of almost 50 minutes. DB2b is a 100Hz downsampled version of DB2a.
  \item DB3~\cite{BCI4Berlin} (BCI Competition IV\footnote{http://bbci.de/competition/iv/}): 59-channel, 1000Hz, 16bps  EEG of 7 subjects performing motor imagery tasks. The database is comprised of 14 recordings of lengths ranging from 29 to 41 minutes, with an average duration of 35 minutes.
  \item DB4~\cite{Delorme2004103}: 31-channel, 1000Hz, 16bps EEG of 15 subjects performing image classification and recognition tasks.
  The database consists of 373 recordings with an average duration of 3.5 minutes, a minimum of 3.3 minutes, and a maximum of 5.5 minutes.
  \item DB5~\cite{PhysioNet, PTB1} (Physikalisch-Technische Bundesanstalt (PTB) Diagnostic ECG Database):  standard 12-lead, 1000Hz, 16bps ECG. This database consists of 549 recordings taken from 290 subjects, with an average duration of 1.8 minutes, a minimum of 0.5 minutes and a maximum of 2 minutes.
\end{itemize}

For the ECG data we extracted leads $i$, $ii$, $v_1\ldots v_6$, to form an
8-channel signal from the standard 12-lead ECG records~\cite{macfarlane2010comprehensive} (the
remaining 4 leads are linear combinations of these 8 channels). Each of these ECG leads is a linear
combination of electrode measures, which represent heart depolarization along the direction of a
certain vector; the notion of distance between channels that we use to determine a coding tree for
Algorithm~\ref{alg:Code} in this case is the angle between these vectors (see
Section~\ref{sseq:knownPositions}).

\subsection{Evaluation procedure}

For each database, we compress each data file separately, and we calculate
the \emph{compression
  ratio} (CR), in bits per sample, defined as
$\text{CR} = L/N$, where $N$ is the sum of the number of scalar samples
over all files of the database, and $L$ is the sum of the number of bits over all compressed files
of the database. \brev Notice that smaller values of CR correspond to better
compression performance\erev. The above procedure is repeated for
$\delta=0,1,2,\ldots,10$. \brev Each discrete sample reconstructed by the
decoder differs by no more than $\delta$ from its original value, which
translates to a maximum difference between signal samples measured in
microvolts ($\mu V$) that depends on the resolution of the acquisition
system. The scaling factor that maps discrete sample differences to voltage
differences in $\mu V$ is 1 for DB1a and DB1b, 0.1 for DB2a, DB2b and DB3,
approximately 0.84 for DB4,\footnote{The exact value depends on the specific
file and channel. The average value is 0.84 with a standard deviation of
0.043} and  0.5 for DB5.

In the experiments, we set the maximum prediction order
in~(\ref{eq:defPrediction}) to $P=7$, the exponential decay factor
in~(\ref{eq:totalErrorP}) to $\lambda=0.99$, and the constant $c$
in~(\ref{eq:defweight}) to a baseline value $c=32$ (in fact, to improve
numerical stability in~(\ref{eq:defPrediction}), we found it useful to
increment [decrement] $c$ whenever the normalization factor $M$
falls below [above] a certain threshold). For Golomb codes,
we set the upper bound on code word length, $\maxCodeWordLength$, to 4 times
the number of bits per sample
of the original signal, and the interval between halvings of statistics
to $\GolombForget = 16$. For each database, we executed
algorithm~\ref{alg:Code}
with $\delta=0$ and all possible choices of a root channel $r$ of the coding tree; the difference between the best and worst root choices was
always less than $0.01$bps. All the results reported in the sequel were
obtained, for each database, with the root channel that yielded the median
compression ratio for that database\erev.

\newcommand {\otoprule }{\midrule [\heavyrulewidth ]}
\begin{table*}
    \caption{\label{tab:CR}Compression ratio of Algorithm~\ref{alg:Code} and
    best compression ratio in~\cite{srinivasan13bhi,dauwels13bhi,ALS} (in
    parenthesis).}
    \centering %
    \setlength{\tabcolsep}{12pt}
    \scriptsize
    \begin{tabular}{rrrrrrrr}
        \toprule
        $\delta$            & DB1a  &        DB1b    &            DB2a
        &                DB2b &     DB3              &    DB4 &    DB5\\
        \otoprule
        0 & 	(5.37) \textbf{4.70} & 	(5.45) \textbf{4.79} & 	(5.69)
        \textbf{5.21} & 	(7.90) \textbf{6.93} & 	(6.46) \textbf{5.42} & 	
        (3.73) \textbf{3.58} & 	(5.03) \textbf{4.78} \\
        5 & 	 \textbf{1.97} & 	 (2.51) \textbf{1.98} & 	
        \textbf{2.34} & 	(4.76) \textbf{3.54} & 	(7.05) \textbf{2.43} & 	
        \textbf{1.79} & 	 \textbf{1.99} \\
        10 & 	 \textbf{1.55} & 	     (1.81) \textbf{1.55} & 	
        \textbf{1.84} & 	(3.85) \textbf{2.73} & 	(6.08) \textbf{1.88} & 	
        \textbf{1.53} & 	 \textbf{1.59} \\
        \bottomrule
    \end{tabular}
\end{table*}

\begin{table*}
\caption{\label{tab:CR2}Compression ratio of Algorithm~\ref{alg:Code2} and percentage relative difference with respect to Algorithm~\ref{alg:Code} (in parenthesis).}
\centering %
\setlength{\tabcolsep}{12pt}
\scriptsize
\begin{tabular}{rrrrrrrr}
\toprule
$\delta$            & DB1a  &        DB1b    &            DB2a &                DB2b &     DB3              &    DB4 &    DB5\\
\otoprule
0 & 	(3.40) 4.54 & 	(3.55) 4.62 & 	(1.15) 5.15 & 	(2.02) 6.79 & 	(0.55) 5.39 & 	(-0.56) 3.60 & 	(-0.21) 4.79 \\
5 & 	(3.05) 1.91 & 	(3.54) 1.91 & 	(2.14) 2.29 & 	(3.95) 3.40 & 	(1.23) 2.40 & 	(-0.56) 1.80 & 	(-1.01) 2.01 \\
10 & 	(1.94) 1.52 & 	(1.94) 1.52 & 	(1.09) 1.82 & 	(4.40) 2.61 & 	(0.53) 1.87 & 	(-0.65) 1.54 & 	(-1.26) 1.61 \\
\bottomrule
\end{tabular}
\end{table*}
\brev

%
%
%
%
%
%
\subsection{Compression Results for Algorithm~\ref{alg:Code}}
\begin{figure}
\centering
  \includegraphics[width=0.45\textwidth]{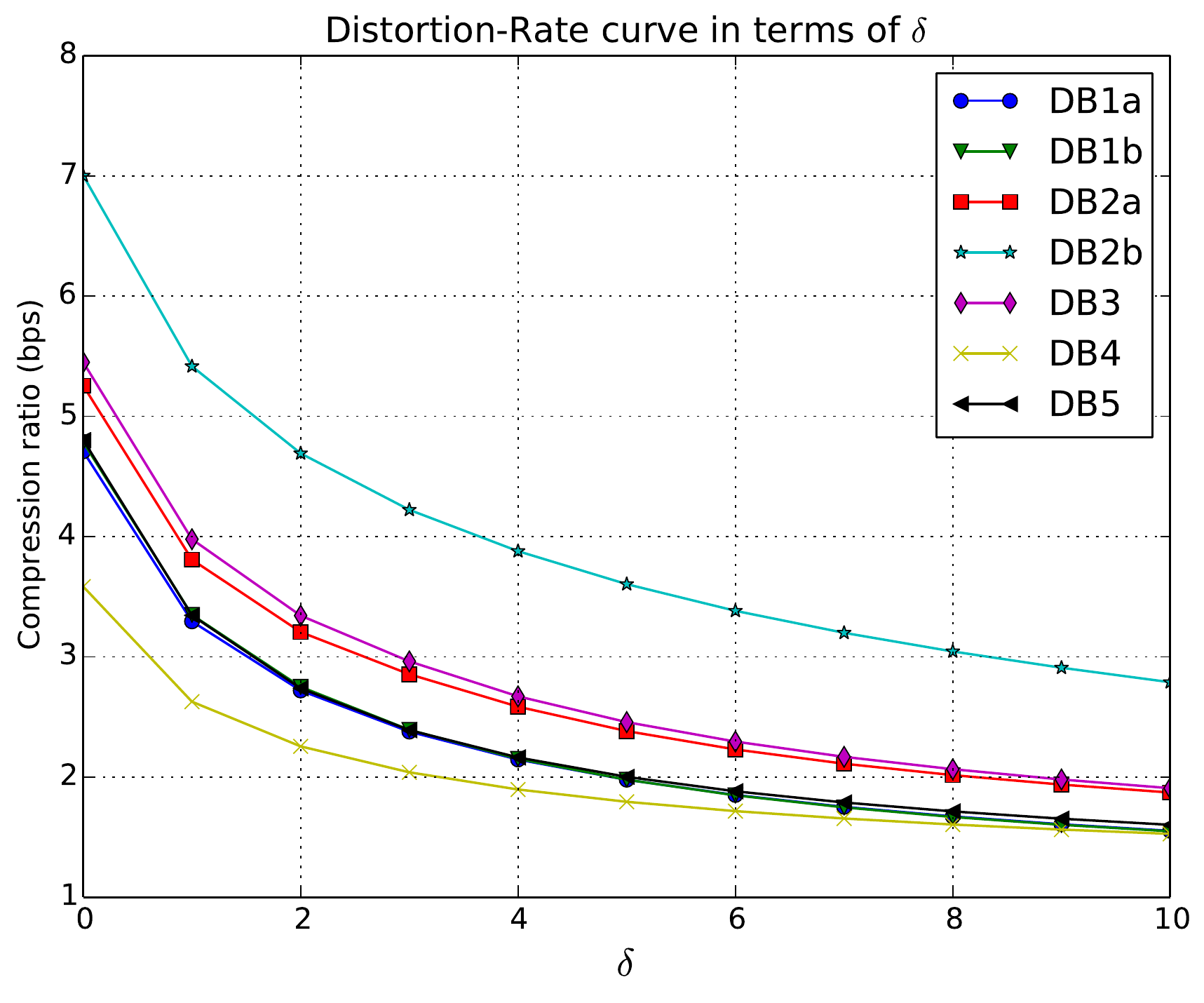}
  \caption{ \label{fig:CR} Compression ratio obtained with Algorithm~\ref{alg:Code} for each value of $\delta$ and all databases. The plots of DB1a, DB1b, and DB5 overlap.}
\end{figure}

The compression ratio obtained with Algorithm~\ref{alg:Code} for each
database, as a function of $\delta$, is shown in Table~\ref{tab:CR} and
plotted in Figure~\ref{fig:CR}.
For $\delta = 0$ (i.e., lossless compression), Table~\ref{tab:CR} also shows, in parenthesis, the compression ratio obtained with the reference software implementation of ALS,\footnote{http://www.nue.tu-berlin.de/menue/forschung/projekte/beendete\textunderscore projekte/ mpeg-4\textunderscore audio\textunderscore lossless\textunderscore coding\textunderscore als} configured for compression rate optimization.
For $\delta > 0$, the value in parenthesis is the best compression reported in~\cite{srinivasan13bhi,dauwels13bhi}, where several compression algorithms are tested with EEG data taken from databases  DB1a, DB2b, and DB3.
As Table~\ref{tab:CR} shows, the compression ratios obtained with
Algorithm~\ref{alg:Code} are the best in all cases.
In the near-lossless mode with $\delta > 0$, the algorithm is designed to
guarantee a worst-case error magnitude of $\delta$ in the reconstruction of
each sample. For completeness, it may also be of interest to assess the
performance of the algorithm under other disortion measures (e.g. mean
absolute error, or SNR), for which it was not originally optimized. Such an
assessment is presented in the Appendix.

\subsection{Compression Results for Algorithm~\ref{alg:Code2}}
\begin{figure}
\centering
  \includegraphics[width=0.45\textwidth]{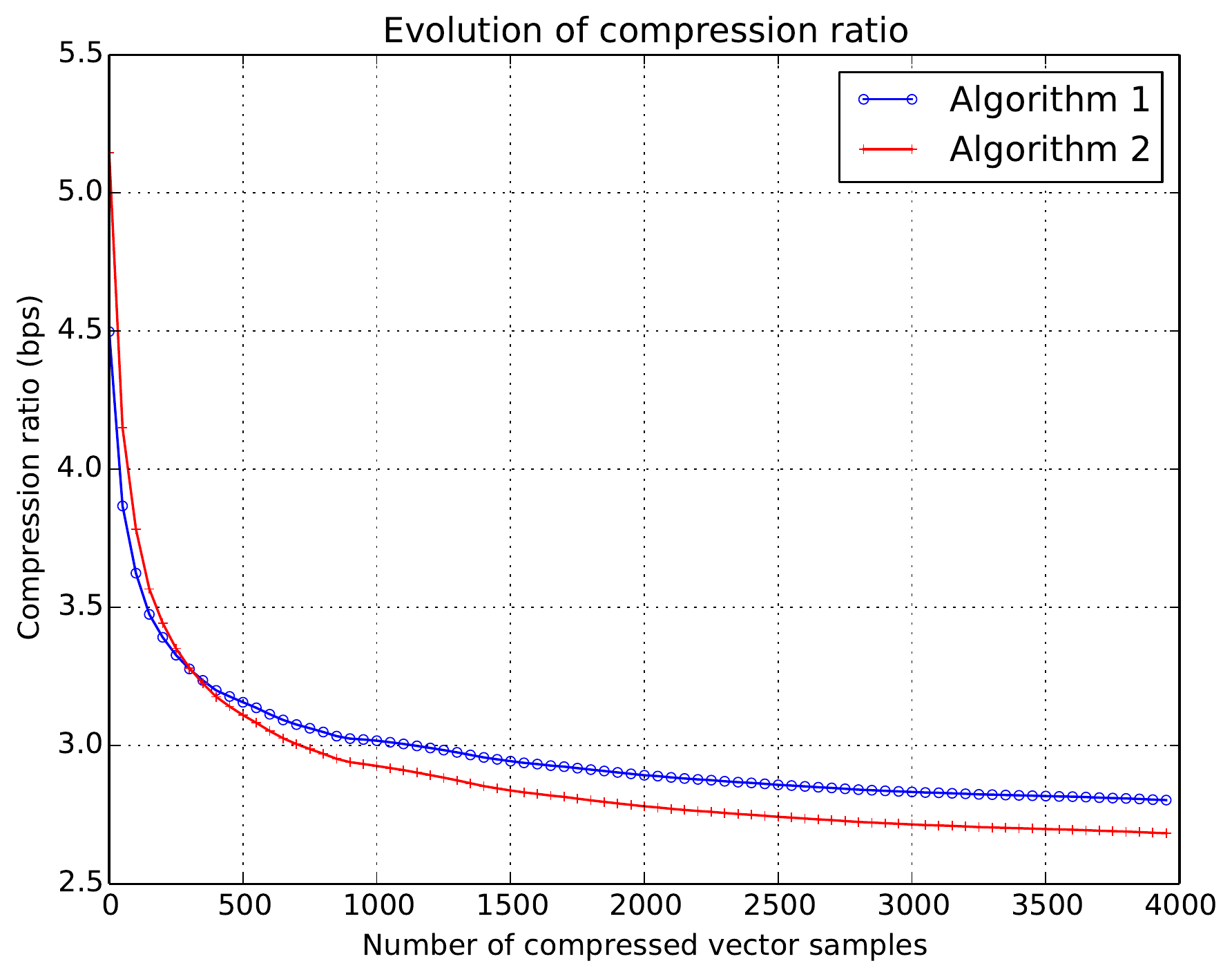}
  \caption{ \label{fig:CRevolDB2b} Evolution in time of the average compression ratio obtained with Algorithm~\ref{alg:Code} and Algorithm~\ref{alg:Code2} for DB2b with $\delta=10$.}
\end{figure}

For Algorithm~\ref{alg:Code2}, we set the block size $B$ equal to $50$, and
for the stopping condition in the coding tree $T$ learning stage, we set
$V=5$, $\gamma = 0.03$, and $N_s = 3000$ (see Section~\ref{ssec:unknown}).
The compression ratio obtained with Algorithm~\ref{alg:Code2}, for each
database and for different values of $\delta$, is shown in
Table~\ref{tab:CR2}. The table also shows, in parenthesis, the percentage
relative difference, $\frac{CR_1-CR_2}{CR_1}\times 100$, between the
compression ratios $CR_1$ and $CR_2$ obtained with Algorithm~\ref{alg:Code}
and Algorithm~\ref{alg:Code2}, respectively, with respect to $CR_1$.

We observe that both algorithms achieve very similar compression ratios in
all cases. In fact, for all the databases except DB4 and DB5,
Algorithm~\ref{alg:Code2} yields better results. Thus, in most of the tested
cases, we obtain better compression ratios by constructing a coding tree from
learned compression statistics rather than based on the fixed geometry, and
this compression ratio improvement is sufficiently large to overcome the cost
incurred during the training segment of the signal, when  a
good coding tree is not yet known. This is graphically illustrated in
Figure~\ref{fig:CRevolDB2b}; for each $n$, in steps of 50 up to a maximum of
4,000 vector samples, the figure plots the compression ratio obtained up to
the encoding of vector sample $n$, averaged over all files of the database
DB2b with $\delta = 10$. For databases DB4 and DB5, although
Algorithm~\ref{alg:Code2} does not surpass Algorithm~\ref{alg:Code}, the
results are extremely close.

Ultimately, which of the algorithms will perform better is a function
of the accuracy of the hypothesis that closer physical proximity of
electrodes implies higher correlation (surely other physical factors
must also affect correlation), and of the heuristic employed to
determine a stopping time for the learning stage of Algorithm~2. The
longer we let the algorithm learn, the higher the likelihood that it
will converge to the best coding tree (which may or may not coincide
with the tree of Algorithm~1), but the higher the computational
cost. The results in Table~\ref{tab:CR2} suggest that the physical
distance hypothesis seems to be more accurate for DB4 and DB5 than for
the other databases, and that the heuristic for $\stopn$ chosen in the
experiments offers a good compromise of computational complexity
against compression performance.

\subsection{Computational complexity}
As mentioned, deriving a coding tree from the signal data in
Algorithm~\ref{alg:Code2} comes at the price of additional memory
requirements and execution time compared to Algorithm~\ref{alg:Code}. These
additional resources are required while the algorithm is learning a
coding tree $T$. The stopping time, $\stopn$, of this learning stage, is
determined adaptively, as specified in~(\ref{eq:stopping}).
Table~\ref{tab:stop} shows the mean and standard deviation of $\stopn$, taken
over the files of each database. We notice that, in general, the update of
$T$ is stopped after a few thousand vector samples. The percentage fraction
of the mean stopping time with respect to the mean total number of vector
samples
in each database is also shown, in parenthesis, in Table~\ref{tab:stop}. We
observe that $n_s$ is less than 1\% of the number of vector samples in most
cases and it is never more than 10\% of that number.

\begin{table*}
\caption{\label{tab:stop}Mean and standard deviation of the stopping time, $n_s$, obtained for each file on different databases. The percentage fraction of the mean stopping time with respect to the mean total number of vector samples in each database is shown in parenthesis.}
\centering %
\setlength{\tabcolsep}{5pt}
\scriptsize
\begin{tabular}{rrrrrrrr}
\toprule
$\delta$            & DB1a  &        DB1b    &            DB2a &                DB2b &     DB3              &    DB4 &    DB5\\
\otoprule
0 & 	(3.49) 637 $\pm$ 154 & 	(6.78) 662 $\pm$ 210 & 	(0.04) 875 $\pm$ 144 & 	(0.3) 713 $\pm$ 222 & 	(0.06) 743 $\pm$ 176 & 	(0.58) 1227 $\pm$ 346 & 	(1.40) 1523 $\pm$ 398\\
5 & 	(4.10) 747 $\pm$ 189 & 	(7.81) 762 $\pm$ 244 & 	(0.04) 1050 $\pm$ 229 & 	(0.44) 1038 $\pm$ 400 & 	(0.09) 1046 $\pm$ 311 & 	(0.58) 1242 $\pm$ 348 & 	(1.76) 1911 $\pm$ 507\\
10 & 	(3.79) 690 $\pm$ 180 & 	(7.35) 717 $\pm$ 236 & 	(0.04) 1044 $\pm$ 220 & 	(0.46) 1081 $\pm$ 439 & 	(0.09) 1075 $\pm$ 339 & 	(0.55) 1172 $\pm$ 327 & 	(1.60) 1735 $\pm$ 439\\
\bottomrule
\end{tabular}
\end{table*}

We measured the total time required to compress and decompress all the files
in all the testing databases. Algorithm~\ref{alg:Code} was implemented in
the  C language, and Algorithm~\ref{alg:Code2} in C++. They were compiled
with GCC 4.8.4, and run
with no multitasking (single thread), on a personal computer with a 3.4GHz
Intel i7 4th generation processor, 16GB of RAM, and under a Linux 3.16 kernel.
Table~\ref{tab:times} shows the average time required by our implementation
of Algorithm~\ref{alg:Code2} to compress and decompress a vector sample for
each database. The results are given for the overall processing of files and
for each main stage of the algorithm separately, namely the first stage
during which the coding tree $T$ is being updated and the second stage during
which $T$ is fixed.

Notice that for a fixed coding tree, the execution time is linear in
the number of scalar samples, with a very small variation from dataset to dataset.
The results that we obtained for Algorithm~\ref{alg:Code} are similar
(smaller in all cases) to those reported for the second stage of
Algorithm~\ref{alg:Code2}. 
On the other hand, for vector samples taken before the stoping time $\stopn$,
the average compression and decompression time is approximately proportional
to $m^2$, as expected. For most databases, this compression time rate exceeds
the sampling rate and, thus, real time compression and decompression would
require buffering data to compensate for this rate difference; notice that
the size required for such a buffer can be estimated from the upper bound
$N_s$ on $\stopn$, and the average compression time for each stage of the
algorithm. In general, with the exception of DB1a and DB1b that are comprised
of short files with a large number of channels, the impact of the additional
computational cost for learning the coding tree is relatively small in
relation to the overall processing time.


\begin{table*}
\begin{center}
\setlength{\tabcolsep}{12pt}
\scriptsize
\caption{\label{tab:times}Average processing time (in microseconds) of a
vector sample for Algorithm 2. Time averages are given for the overall
processing of files and also discriminating the period of time during which
the coding tree is being updated ($n\leq\stopn$) from the period of time in
which it is fixed ($n>\stopn$).}
\begin{tabular}{llrrrrrrr}\toprule
 &   & DB1a  & DB1b  & DB2a  & DB2b & DB3 & DB4   & DB5 \\\hline
$m$ &&        64    & 64    & 118   & 118  & 59   & 31   & 8 \\\hline
$n>\stopn$ & coding   & 63.3 & 	63.3 & 	118.4 & 	119.4 & 	58.6 & 	30.4 & 	7.2 \\
$n>\stopn$ & decoding & 63.6 & 	63.6 & 	118.8 & 	119.2 & 	58.9 & 	30.4 & 	7.6 \\
\hline
$n\leq\stopn$& coding  &  7661.4 & 	7667.1 & 	26799.3 & 	27190.6 & 	6304.7 & 	1202.9 & 	65.9 \\
$n\leq\stopn$& decoding& 7887.5 & 	7889.2 & 	27811.6 & 	28367.1 & 	6516.0 & 	1223.2 & 	66.7 \\
\hline
Global & coding   & 320.7 & 	562.0 & 	130.5 & 	217.6 & 	60.8 & 	37.2 & 	8.1 \\
Global & decoding & 328.7 & 	576.7 & 	131.3 & 	221.6 & 	61.2 & 	37.4 & 	8.5 \\
\bottomrule
\end{tabular}
\end{center}
\end{table*}


Algorithm 1 has also been ported to a low power MSP432 microcontroller running at 48MHz, where it
requires an average of $3.21$ milliseconds to process a single scalar sample, allowing a maximum
real-time processing of 16 channels at 321Hz. In order to fit within the memory of the MSP432, the
maximum order of the predictors was set to $P=4$, resulting in a slight performance degradation
with respect to that reported in Table~\ref{tab:CR}. The resulting memory footprint is $42.5$kB
of flash memory, and $26.7$kB of RAM. Further details and advances in this
direction will be published elsewhere.

\erev

\section{Conclusions}
\label{sec:concl}
The space and time redundancy in both EEG and ECG can be effectively reduced 
with a predictive adaptive coding scheme in which each channel is predicted 
through an adaptive linear combination of past samples from the same channel, 
together with past and present samples from a designated reference channel. 
Selecting this reference channel according to the physical distance between 
the sensors that register the signals is a very simple approach, which yields 
good compression rates and, since it can be implemented off-line, incurs no 
additional computational effort at coding or decoding time. When sensor 
positions are not known a priori, we propose a scheme for inter-channel 
redundancy analysis based on efficient minimum spanning tree construction 
algorithms for directed graphs. Both proposed encoding algorithms are 
sequential, and thus suitable for low-latency applications. In addition, the 
number of operations per time sample, and the memory requirements depend 
solely on the number of channels of the acquisition system, which makes the 
proposed algorithms attractive for hardware implementations. \brev This is 
especially true for Algorithm~\ref{alg:Code}, whose memory and time 
requirements depend linearly on the number of channels. In the case of 
Algorithm~\ref{alg:Code2} these requirements are quadratic in the number of 
channels during the learning stage, which might make a pure hardware 
implementation more difficult if this number is large\erev.

\appendix[Other distortion measures]
\brev
\begin{table*}
\caption{\label{tab:MAE}MAE (in $\mu V$) obtained with
Algorithm~\ref{alg:Code} and  Algorithm~\ref{alg:Code2} (in parenthesis) for
different values of $\delta$.}
\centering %
\setlength{\tabcolsep}{12pt}
\scriptsize
\begin{tabular}{rrrrrrrr}
\toprule
$\delta$            & DB1a  &        DB1b    &            DB2a &                DB2b &     DB3              &    DB4 &    DB5\\
\otoprule
5 & ( 2.69)  2.69 & ( 2.70)  2.70 & ( 0.27)  0.27 & ( 0.27)  0.27 & ( 0.27)  0.27 & ( 2.30)  2.29 & ( 1.36)  1.36\\
10 & ( 5.00)  5.00 & ( 5.08)  5.09 & ( 0.52)  0.52 & ( 0.52)  0.52 & ( 0.52)  0.52 & ( 4.39)  4.38 & ( 2.59)  2.58\\
\bottomrule
\end{tabular}
\end{table*}

\begin{table*}
\caption{\label{tab:SNR}SNR (in dB) obtained with Algorithm~\ref{alg:Code}
and  Algorithm~\ref{alg:Code2} (in parenthesis) for different values of
$\delta$.}
\centering %
\setlength{\tabcolsep}{12pt}
\scriptsize
\begin{tabular}{rrrrrrrr}
\toprule
$\delta$            & DB1a  &        DB1b    &            DB2a &                DB2b &     DB3              &    DB4 &    DB5\\
\otoprule
5 & (27.79) 27.59 & (27.40) 26.98 & (49.47) 49.47 & (49.48) 49.47 & (48.11) 48.11 & (59.70) 59.69 & (53.07) 52.89\\
10 & (22.35) 22.16 & (21.87) 21.46 & (43.83) 43.83 & (43.84) 43.83 & (42.52) 42.52 & (54.07) 54.09 & (47.48) 47.33\\
\bottomrule
\end{tabular}
\end{table*}

\begin{table*}
\caption{\label{tab:SNR2}SNR (in dB) obtained with Algorithm~\ref{alg:Code}
using a value of $\delta$ that corresponds approximately to 1 $\mu V$ for
each database. The specific value of $\delta$ used in each case is shown in
parenthesis.}
\centering %
\setlength{\tabcolsep}{12pt}
\scriptsize
\begin{tabular}{rrrrrrr}
\toprule
DB1a  &        DB1b    &            DB2a &                DB2b &     DB3              &    DB4 &    DB5\\
\otoprule
 (1) 39.26 & (1) 38.69 & (10) 43.83 & (10) 43.83 & (10) 42.52  & (1) 71.46 & (2) 59.87\\
\bottomrule
\end{tabular}
\end{table*}

\begin{figure}
\centering
  \includegraphics[width=0.45\textwidth]{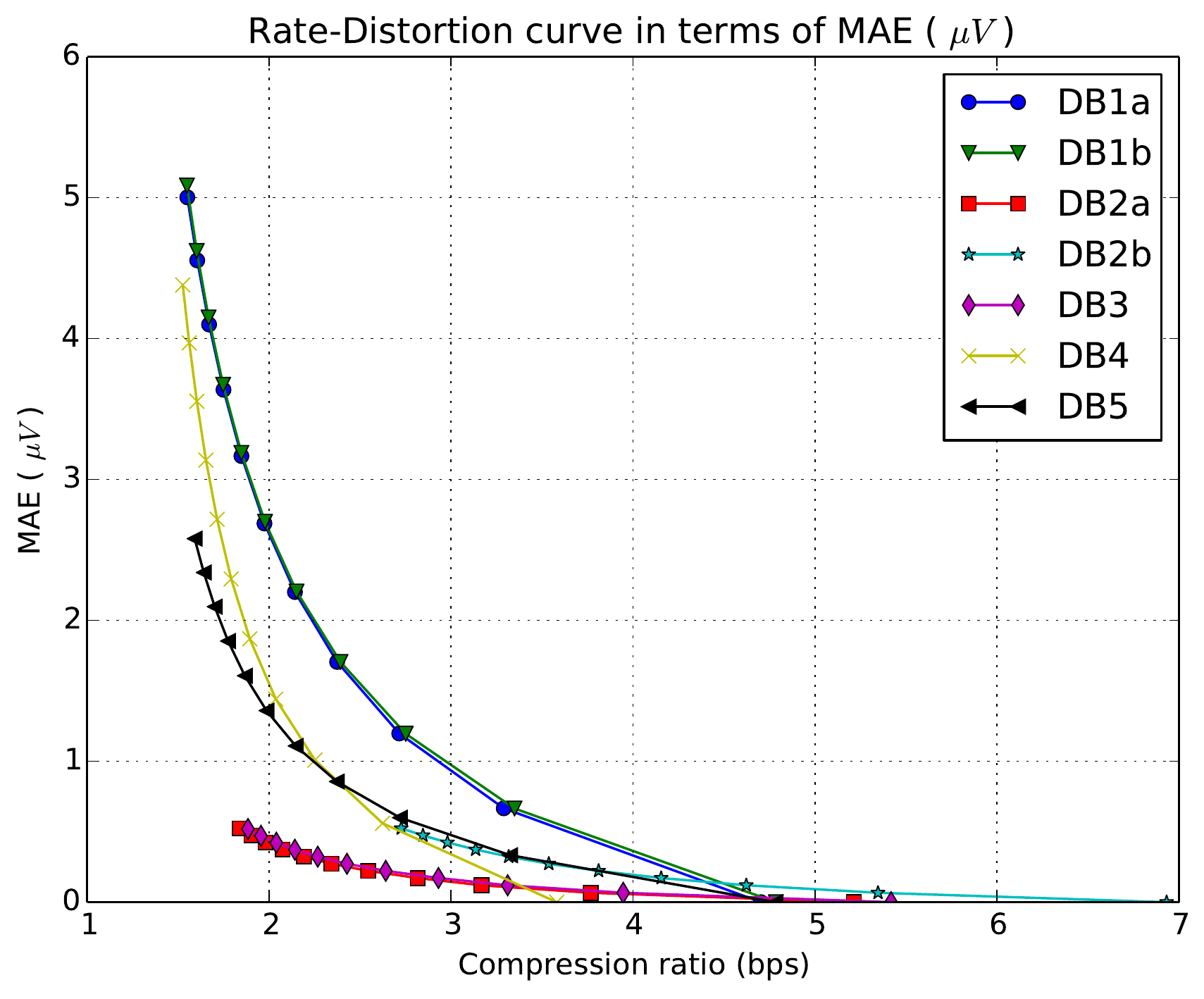}
  \caption{ \label{fig:MAE} Rate-Distortion curve in terms of MAE obtained with Algorithm~\ref{alg:Code} when $\delta$ takes the values $0, 1, \ldots, 10$ for all databases.}
\end{figure}

\begin{figure}
\centering
  \includegraphics[width=0.45\textwidth]{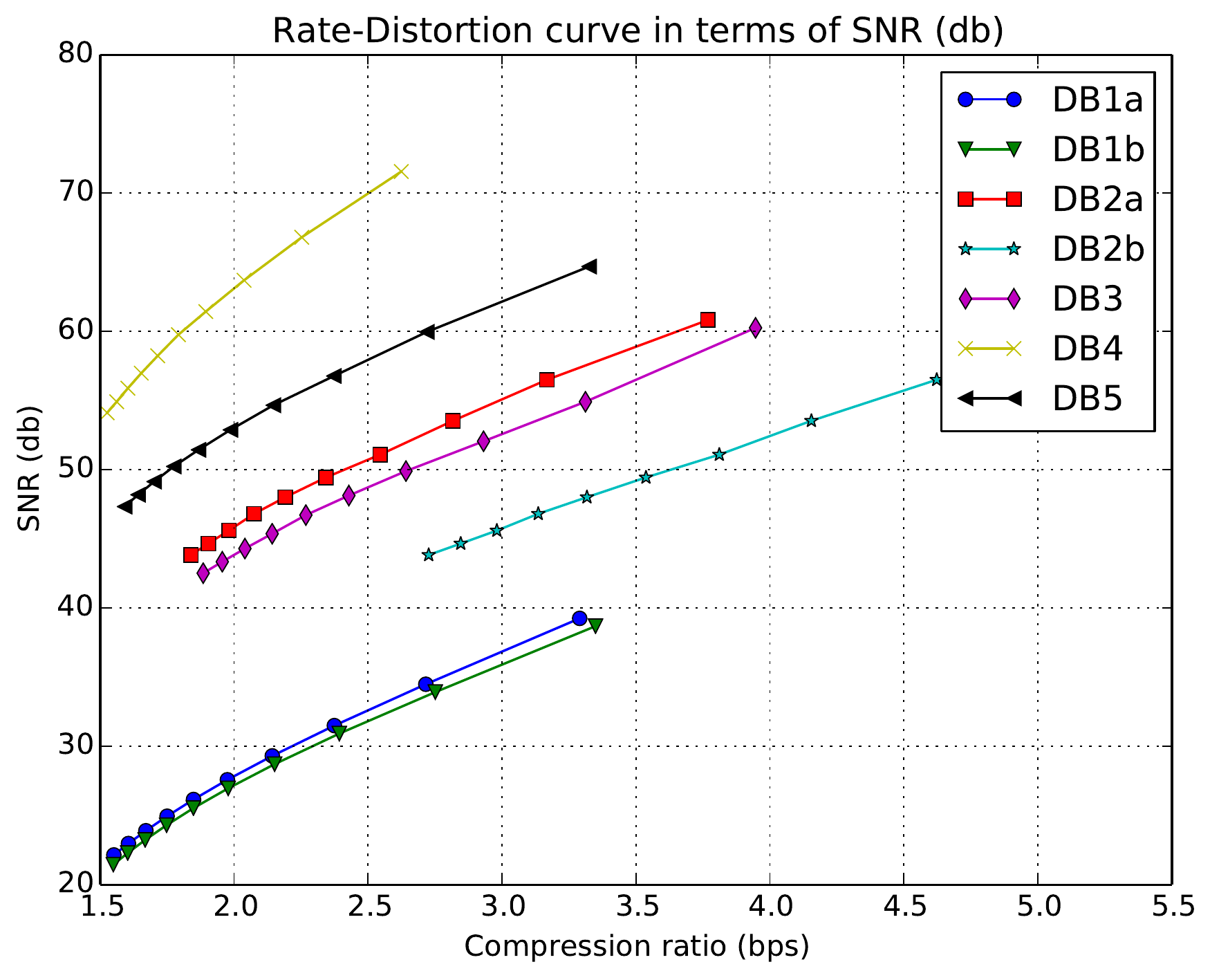}
  \caption{ \label{fig:SNR} Rate-Distortion curve in terms of SNR obtained with Algorithm~\ref{alg:Code} when $\delta$ takes the values $1, 2, \ldots, 10$ for all databases.}
\end{figure}

For the cases where
$\delta > 0$ (near-lossless) we compute the \emph{mean absolute error}
(MAE), and the \emph{signal to noise ratio} (SNR)
given respectively by
\begin{align*}
  \mathrm{MAE} &= \frac{1}{Nm}\sum_{i=1}^{m}\sum_{n=1}^N|\sample{i}{n}-\qsample{i}{n}|\,,\\
  \mathrm{SNR} &=
  10\log_{10}\frac{\sum_{i=1}^{m}\sum_{n=1}^N\sample{i}{n}^2}
  {\sum_{i=1}^{m}\sum_{n=1}^N(\sample{i}{n}-\qsample{i}{n})^2}\,.
\end{align*}
Figures~\ref{fig:MAE} and~\ref{fig:SNR} show plots of the MAE and the
SNR, respectively, against the compression ratio obtained with
Algorithm~1 for different values of $\delta$. The plots for
Algorithm~2 are very similar and thus omitted. The MAE and SNR
obtained with both algorithms for $\delta = 5$ and $\delta = 10$ are
shown in Tables~\ref{tab:MAE} and~\ref{tab:SNR}, respectively. We
observe that, in every case, the MAE is close to one half of the
maximum allowed distortion given by $\delta$ (appropriately scaled to
$\mu V$). This matches well the behavior expected from a TSGD
hypothesis on the prediction errors. The SNR varies significantly
among databases for a fixed value of $\delta$, due to both the
difference in scale and the difference in power (in $\mu V^2$) among
the databases.  Table~\ref{tab:SNR2} shows the SNR obtained using a
value of $\delta$ that corresponds approximately to 1 $\mu V$ for each
database. We still observe very different values of SNR, which is
explained by the difference in power of the
signals.  
We verified by direct observation that, as expected by design, the
\emph{maximum absolute error} (M*AE) for Algorithm~\ref{alg:Code} is
equal to $\delta$ in every case.

Table~\ref{tab:distortion} compares the SNR and M*AE of
Algorithm~\ref{alg:Code} with the results published
in~\cite{dauwels13bhi}. For the comparison, we selected the value of
$\delta$ that yields the compression ratio closest to that reported
in~\cite{dauwels13bhi} in each case. We observe that the SNR is in
general better for the best algorithm in~\cite{dauwels13bhi}, except
in the case of DB3, where Algorithm~\ref{alg:Code} attains similar (or
better) compression ratios with \emph{lossless} compression (infinite
SNR). The M*AE is much smaller for Algorithm~\ref{alg:Code} in all
cases.  {These results should be taken with a grain of salt, though,
  given that our scheme, contrary to that of~\cite{dauwels13bhi}, does
  not target SNR.
}

\begin{table*}
  \caption{\label{tab:distortion}Best compression ratio / distortion reported in~\cite{dauwels13bhi} (in parenthesis), and distortion obtained with  Algorithm~\ref{alg:Code} for the choice of $\delta$ that yields the closest compression ratio.}
\centering %
\setlength{\tabcolsep}{12pt}
\scriptsize
\begin{tabular}{lrrr}
\toprule
Database            & CR (bps)  &        SNR (db)    &            M*AE ($\mu V$) \\
\otoprule
DB1b & 	(3.32) 3.35 & 	(\textbf{47.3}) 38.7 & 	     (2.85) \textbf{1} \\
DB1b & 	(2.42) 2.39 & 	(\textbf{36.1}) 30.9 & 	     (5.35) \textbf{3} \\
\hline
DB2b & 	(5.26) 5.35 & 	(\textbf{80.0}) 60.8 & 	     (0.73) \textbf{0.1} \\
DB2b & 	(4.29) 4.15 & 	(\textbf{73.9}) 53.5 & 	     (1.22) \textbf{0.3} \\
\hline
DB3 & 	(6.27) 5.42 & 	(80.0) $\mathbf{\infty}$ & 	 (0.67) \textbf{0} \\
DB3 & 	(5.33) 5.42 & 	(66.0) $\mathbf{\infty}$ & 	 (1.19) \textbf{0} \\
\bottomrule
\end{tabular}
\end{table*}



\begin{figure*}[t]
  \centering
  \begin{subfigure}[b]{0.3\textwidth}
    \includegraphics[width=\textwidth]{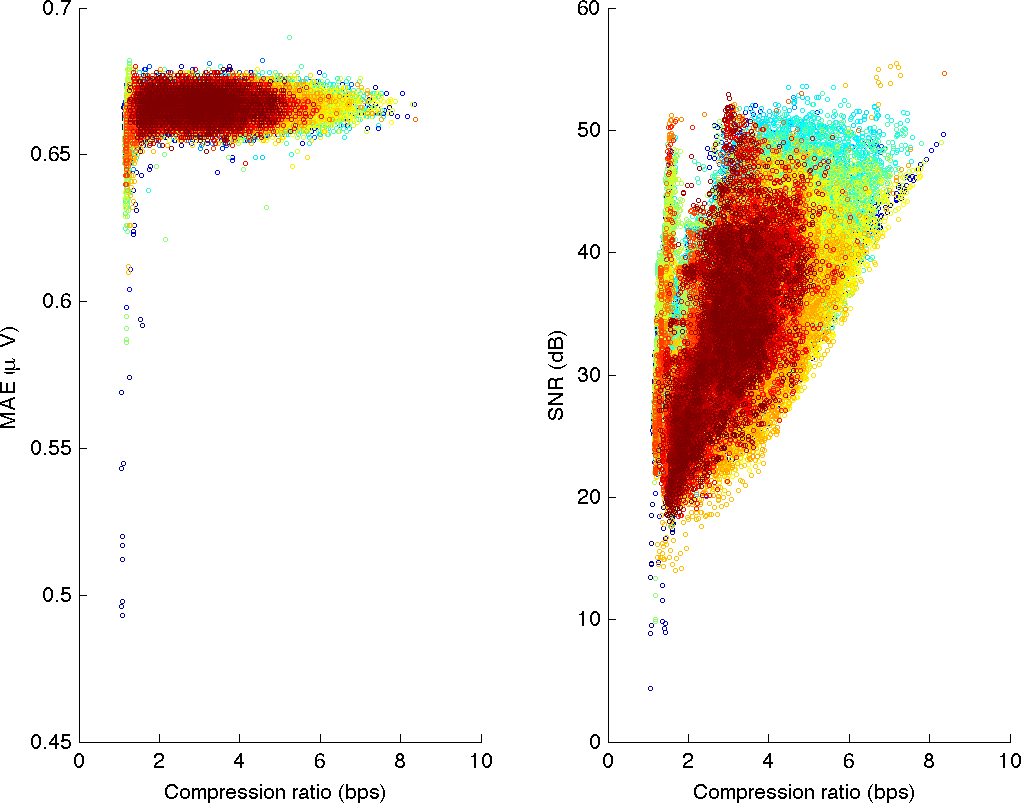}
    \caption{}
  \end{subfigure}\qquad
  \begin{subfigure}[b]{0.3\textwidth}
    \includegraphics[width=\textwidth]{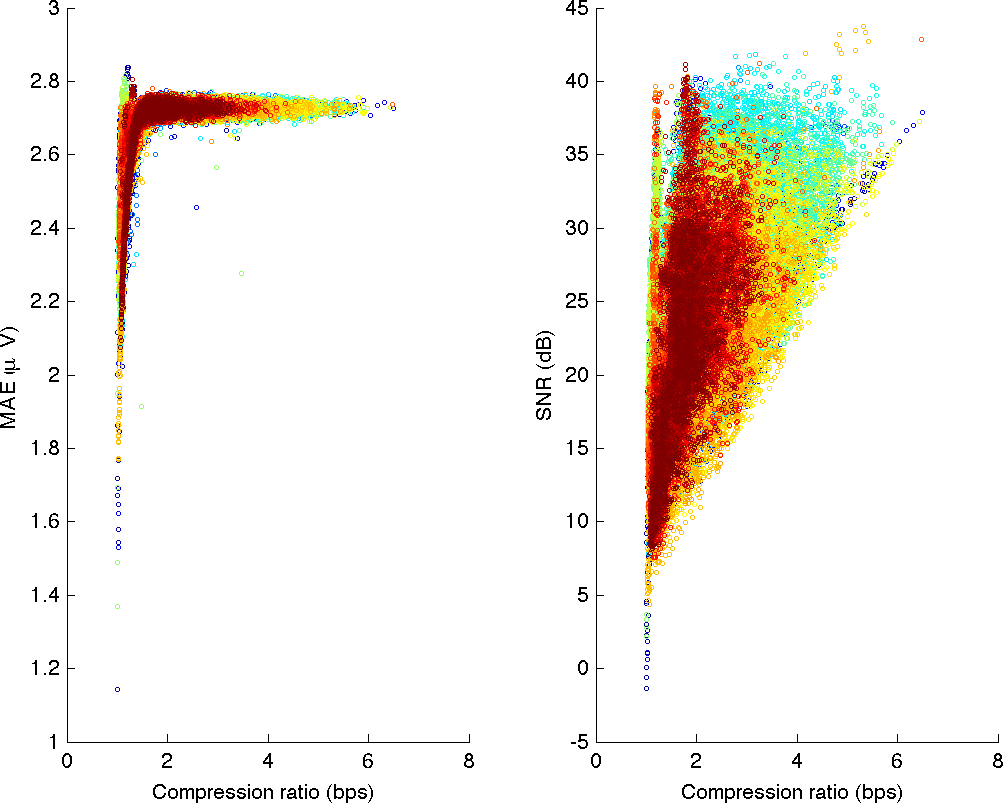}\qquad
    \caption{}
  \end{subfigure}\qquad
  \begin{subfigure}[b]{0.3\textwidth}
    \includegraphics[width=\textwidth]{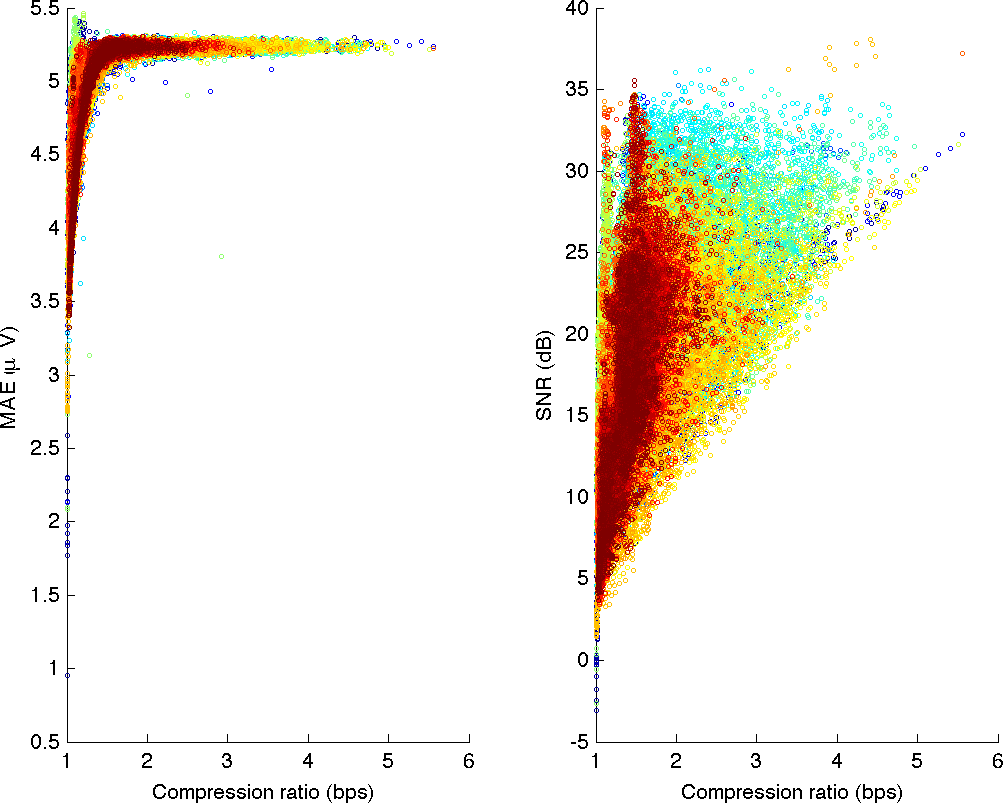}
    \caption{}
  \end{subfigure}
  \caption{\label{fig:MAE+SNR_DB1a} (a)--(c) Rate-Distortion values
    for all channels and files in terms of MAE and SNR obtained with
    Algorithm~\ref{alg:Code} when $\delta$ takes the values (a)~$1$,
    (b)~$5$ and (c)~$10$ for database DB1a. Each channel is plotted
    with a different color. 
      }
\end{figure*}

We also analyze the variation of the MAE and SNR over the channels for
all the databases and its dependence on
$\delta$. Figures~\ref{fig:MAE+SNR_DB1a}(a)-(c) show the MAE and SNR
against the compression ratio for all channels with $\delta =
\{1,5,10\}$ for the database DB1a. All channels have a similar
behavior for the MAE and SNR, with lower dispersion for the MAE than
the SNR.
Since the mean square error has a dispersion similar to the MAE (not
shown), the greater dispersion of the SNR is explained by the
variation of the power among channels. This dispersion can be reduced
by selecting a specific value of $\delta$ for each channel, depending
on the power of its signal. The behavior for MAE and SNR is similar in
all the databases, and, thus, it is reported only for DB1a for
succinctness.

\begin{figure}
\centering
  \includegraphics[width=0.43\textwidth]{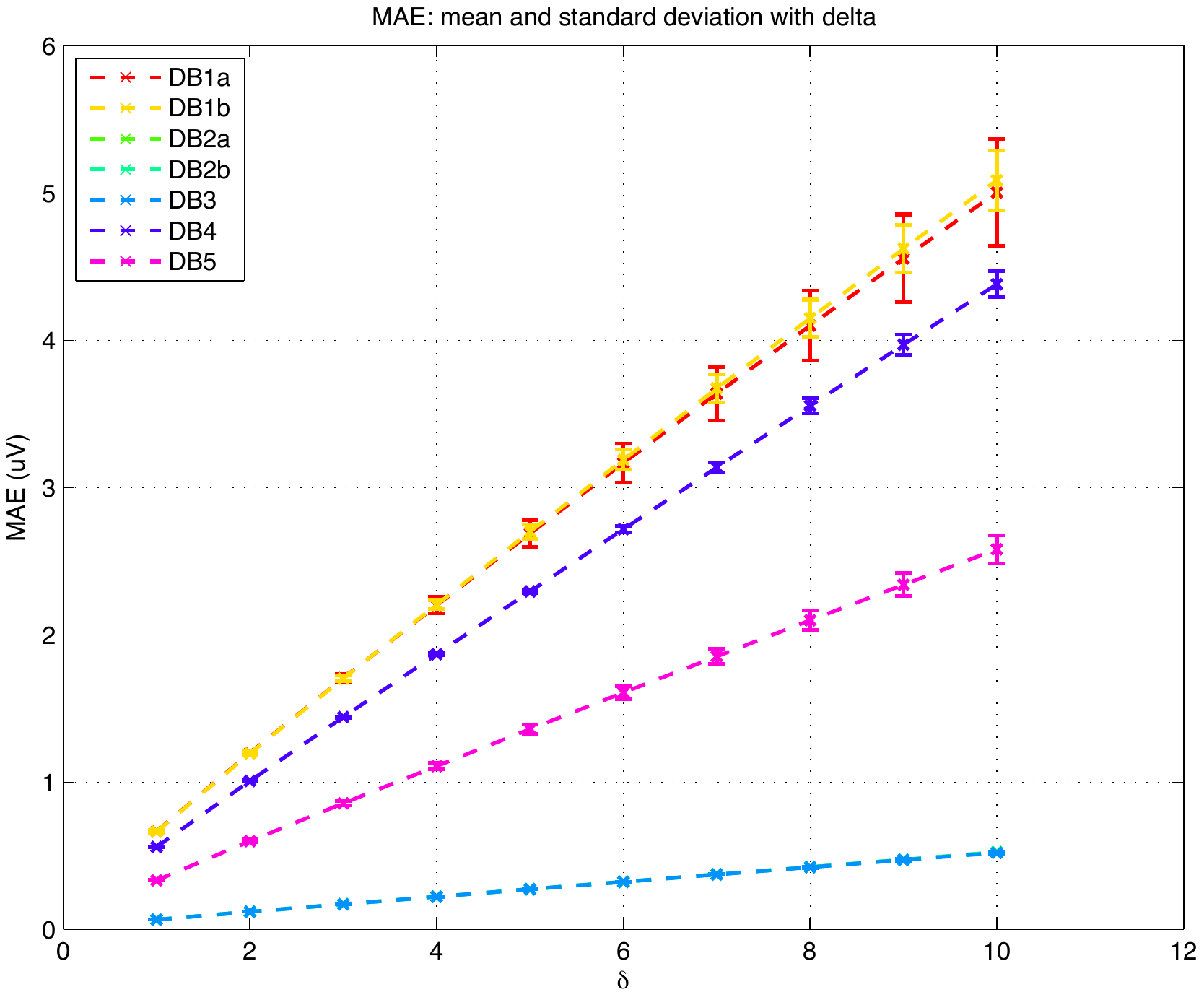}
  \caption{\label{fig:MAEmean} MAE mean over all channels and its
    standard deviation as an errorbar when $\delta$ takes the values
    $1, 2,\dots, 10$ for all databases.}
\end{figure}

\begin{figure}
\centering
  \includegraphics[width=0.43\textwidth]{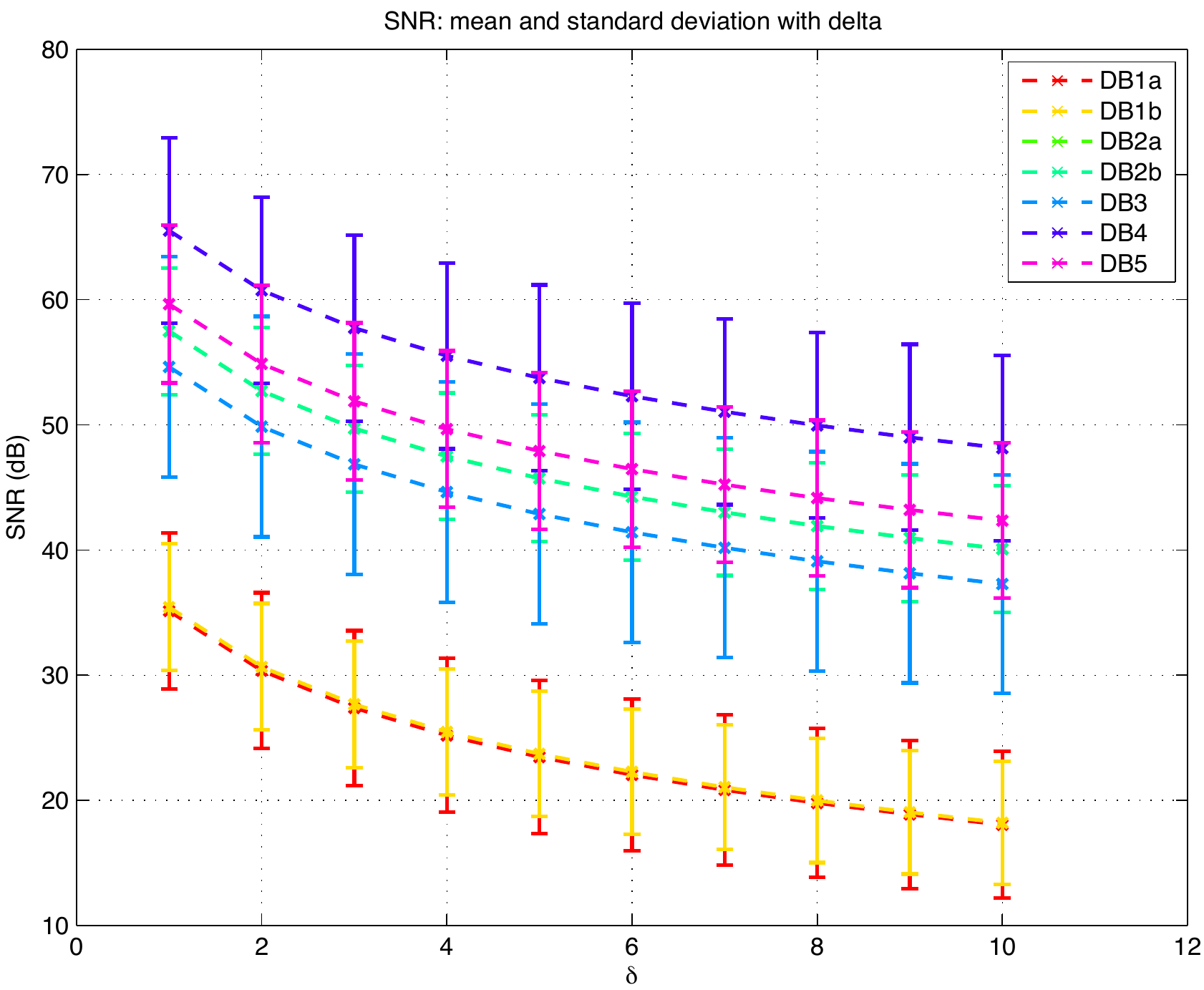}
  \caption{\label{fig:SNRmean} SNR mean over all channels and its
    standard deviation as an errorbar when $\delta$ takes the values
    $1, 2,\dots, 10$ for all databases.}
\end{figure}

Figures \ref{fig:MAEmean} and \ref{fig:SNRmean} summarize the mean and standard deviation of both the MAE and
the SNR measures over all channels, for each database
for different values of $\delta$. One standard deviation is shown as
an errorbar with the mean for each $\delta$. Again, a similar behavior
among all the databases is observed, with lower dispersion in the MAE,
increasing with $\delta$ and almost constant dispersion for the SNR.

\erev

\balance
\bibliographystyle{IEEEtran}
\bibliography{references}

\end{document}